\documentclass{aa}
\usepackage{amsmath,graphics,amssymb}
\usepackage[varg]{txfonts}
\usepackage{natbib}
\usepackage{times}
\usepackage{frcursive}

\newfont{\eufont}{eufm10}

\newcommand{\dmdt}{\dot{M}}

\newcommand{\zav}[1]{\left(#1\right)}
\newcommand{\hzav}[1]{\left[#1\right]}
\newcommand{\szav}[1]{\left\{#1\right\}}

\newcommand{\pd}[2]{\frac{\partial #1}{\partial #2}}

\newcommand{\kms}{\ensuremath{\text{km}\,\text{s}^{-1}}}
\newcommand{\vel}{{v}}
\newcommand{\vinfty}{\ensuremath{\vel_\infty}}
\newcommand{\vesc}{\ensuremath{\vel_\text{esc}}}
\newcommand{\ms}{\ensuremath{{M}_{\odot}}}
\newcommand{\msr}{\ensuremath{\ms\,\text{yr}^{-1}}}

\newcommand{\de}{\mathrm{d}}
\newcommand{\Teff}{\mbox{$T_\mathrm{eff}$}}

\newcommand\cc{\ensuremath{C_\text{c}}}

\begin{document}

\title{New mass-loss rates of B supergiants from global wind models}

\author{J.~Krti\v{c}ka\inst{1} \and J.~Kub\'at\inst{2} \and
        I.~Krti\v ckov\'a\inst{1}}

\institute{Department of Theoretical Physics and Astrophysics, Faculty of
           Science, Masaryk University, CZ-611 37 Brno,
           Czech Republic \and
           Astronomical Institute, Czech Academy of Sciences%
\thanks{Astronomick\'y \'ustav, Akademie v\v{e}d \v{C}esk\'e republiky},
           CZ-251 65 Ond\v{r}ejov, Czech Republic%
}

\date{Received}

\abstract{Massive stars lose a significant fraction of mass during their
evolution. However, the corresponding mass-loss rates are rather uncertain,
especially for evolved stars. To improve this, we calculated global line-driven
wind models for Galactic B supergiants. Our models predict radial wind structure
and particularly the mass-loss rates and terminal velocities directly from basic
stellar parameters. The hydrodynamic structure of the flow is
consistently determined from the photosphere in nearly hydrostatic equilibrium to
supersonically expanding wind. The radiative force is derived from the solution
of the radiative transfer equation in the comoving frame. We provide a simple
formula that predicts theoretical mass-loss rates as a function of stellar
luminosity and effective temperature. The mass-loss rate of B supergiants slightly
decreases with temperature down to about 22.5\,kK, where the region of
recombination of \ion{Fe}{iv} to \ion{Fe}{iii} starts to appear. In this region,
which is about 5\,kK wide, the mass-loss rate gradually increases by a factor of
about 6. The increase of the mass-loss rate is associated with a gradual
decrease of terminal velocities by a factor of about 2. We compared the
predicted wind parameters with observations. While the observed wind terminal
velocities are reasonably reproduced by the models, the situation with mass-loss
rates is less clear. The mass-loss rates derived from observations that are
uncorrected for clumping are by a factor of 3 to 9 higher than our
predictions on cool and hot sides of the studied sample, respectively. These
observations can be reconciled with theory assuming a temperature-dependent
clumping factor that is decreasing toward lower effective temperatures. On the other
hand, the mass-loss rate estimates that are not sensitive to clumping agree with
our predictions much better. Our predictions are by a factor of about 10 lower
than the values currently used in evolutionary models appealing for
reconsideration of the role of winds in the stellar evolution.}


\maketitle
%________________________________________________________________

\section{Introduction}

One of the most important challenges of stellar astrophysics is to determine the
fate of single stars as a function of initial stellar parameters. While for low-
and intermediate-mass stars ($M\lesssim8\,M_\odot$) the final remnant is a white
dwarf \citep{vasiwo,milbe}, the endpoint of evolution of massive stars is either
a neutron star or a black hole depending on the initial parameters and evolutionary
history of a star \citep{explomas,pejto}.

Mass loss is one of the most important processes that affects the stellar
evolution and mass of final stellar remnants. The source of the first ever
detected gravitational wave emission may serve as a striking example
\citep{jinyabbott}. As a result of mass loss, a typical mass of a black hole in
our Galaxy is on the order of a solar mass \citep[e.g.,][]{torca,synjunga}, but
the mass of each detected merging black hole was significantly higher, of about
$30\,{M}_\odot$. It is likely that the mass loss of binary black hole
progenitors was partially suppressed, either as a result of low metallicity or
by magnetic field quenching \citep{petvln}.

The importance of mass loss goes beyond its influence on the stellar evolution.
The mass-loss rate determines the dynamical effect of massive stars on
interstellar medium \citep{casbublina} and the amount of elements distributed
across the galaxies \citep{cinskasmrt}. Wind blanketing modifies the number
of ionizing photons produced by massive stars \citep{acko}.

Despite the progress in understanding mass loss from massive stars
\citep[for a review]{pulvina} and its importance, the exact rates are still
rather uncertain. From a theoretical point of view, this is connected with
extreme intricacies of reliable hot star wind models. First of all, mass-loss
rate predictions require advanced global models that calculate the radiative
force using detailed radiative transfer \citep{grahamz,powrdyn,cmfkont,sundyn}.
Moreover, hydrodynamical models predict strong instability connected with
radiative driving \citep{ocr,felto}. For strong base perturbations, the
inhomogeneities already appear at the base of the wind \citep{felpulpal}. This
further modifies theoretically predicted mass-loss rates \citep{muij,irchuch}.

Neither observational indicator provides a reliable way to determine the
mass-loss rates. There are several methods to determine 
%KrEd: how
%
the mass-loss rates
%KrEd: can be derived
%
from observations, but each observational characteristic is in its
specific way influenced by small-scale inhomogeneities (clumping). The shape of
X-ray line profiles is affected by absorption in the cool wind. Therefore, X-ray
line profiles provide a measure of the wind mass-loss rate \citep{cohcar}, which
is not expected to be significantly influenced by optically thin clumps
\citep{lojza,irchuch}. However, the shape of X-ray line profiles may be affected
by clumps that are optically thick \citep{lidarikala}.

The strength of the ultraviolet wind P~Cygni line profiles, which are also
frequently used to determine the mass-loss rate, is also affected by optically
thick and optically thin inhomogeneities \citep{chuchcar,sund,clres1}. However,
a simultaneous fit of several wind lines provides a way to independently
determine both mass-loss rates and clumping parameters \citep{clres2}. These may
be combined with the rates determined from H$\alpha$ line and from radio
emission, which are also severely affected by clumping \citep{pulchuch}. In such
a situation, alternative mass-loss rate determinations based on the interaction of
the wind with the circumstellar environment \citep{henluk,kobul} may provide
estimates that are free of the influence of inhomogeneities.

The realm of B supergiants is affected by uncertainties of mass-loss rates even
more, partly because of a lack of observational estimates that account for
clumping. Although the B supergiants represent a relatively short-lived
evolutionary stage, the mass loss during the B supergiant phase may have
a significant impact on the evolution of mass and angular momentum of massive
stars \citep{kostel}. Massive stars typically lose more mass in later
evolutionary phases than during early stages as O-type stars
\citep[who compiled mass-loss rate predictions from several sources]{gromek}.
The domain of B supergiants is important for the evolution of core-collapse
supernova progenitors. While B supergiants that are direct descendants of
main-sequence stars are not expected to be immediate progenitors of supernovae,
most stars that have exploded as core-collapse supernova also
%KrEd: become B
had been
supergiants
%KrEd: sooner or later,
in the past,
and more massive stars may pass the evolutionary
stage of B supergiants even twice \citep{grohsnpred,sagem}. To improve the
situation with B supergiant mass-loss rate estimates, we provide global wind
models of these stars to consistently predict the wind properties.

\section{Wind models}
\label{winmo}

To model B supergiants, we used the METUJE wind models described by
\citet{cmfkont}. The models were calculated with the following basic
assumptions:
%KrEd: \begin{itemize}
%KrEd: \item
a)
We assumed a spherically symmetric and stationary stellar wind.
%KrEd: \item
b)
The models self-consistently solved the same equations
(continuity equation, equation of motion, energy equation, radiative transfer
equation, and kinetic equilibrium equations) in the photosphere and
in the wind (global models).
%KrEd: \item
c)
We solved radiative transfer in the comoving frame \citep[CMF;][]{mikuh};
%KrEd: \item
d)
The atomic level occupation numbers were derived from the kinetic equilibrium
equations \citep[Chapter~9]{hubenymihalas} with bound-free terms calculated
from the CMF radiative field and bound-bound terms with the Sobolev
approximation \citep{klecany}.
%KrEd: \item
e)
Atomic data for the solution of kinetic equilibrium equations was adopted
mostly from the TLUSTY models \citep{bstar2006} with some updates from the
Opacity and Iron Project data \citep{topt,zel0}.
%KrEd: \item
f)
The models account for the most abundant elements given in \citet{btvit},
assuming solar chemical composition after \citet{asp09}.
%KrEd: \item
g)
The wind density was derived from continuum equation.
%KrEd: \item
h)
The wind velocity was determined from the equation of motion with the
radiative force due to continuum and line transitions.
%KrEd: \item
i)
The temperature was derived from the
radiative equilibrium equation either in integral
or differential form in the photosphere \citep{kubii}, while the model
calculates the radiative heating/cooling using the thermal balance of electrons
method \citep{kpp}.
%KrEd: \item
j)
The wind density, temperature, and velocity were determined simultaneously
using Newton-Raphson method and iterations.
%KrEd: \end{itemize}
We used the TLUSTY plane-parallel static model atmospheres
\citep{ostar2003,bstar2006} to derive the initial estimate of the photospheric
structure.

The models were calculated for a grid of stellar temperatures $\Teff
=10\,000-27\,500\,$K covering the spectral range of B supergiants for three
values of the stellar luminosity $L$ (or mass $M$). The parameters given in
Table~\ref{bvele} (supplemented by stellar radii $R_{*}$ and Eddington
parameters\footnote{The ratio of the radiative acceleration from electron
scattering to the stellar gravity \citep[the quantity $1-\beta$ in][]{Eddlum}.}
$\Gamma$) correspond to typical observational values given in \citet{crow}.

\begin{table}[t]
\caption{Stellar parameters of the model grid with derived values of the
terminal velocity $v_\infty$ and the mass-loss rate $\dot M$.}
\centering
\label{bvele}
\begin{tabular}{ccccc}
\hline
\hline
Model &$\Teff$ & $R_{*}$ & $\vinfty$ & $\dot M$  \\
& $[\text{K}]$ & $[{R}_{\odot}]$ & [\kms] & [\msr] \\
\hline
\multicolumn{5}{c}{$M=25\,{M}_{\odot}$, $\log(L/L_\odot)=5.28$,
$\Gamma=0.18$}\\
275-25 & 27500 & 19.3 & 1890 & $9.1\times10^{-8}$ \\
250-25 & 25000 & 23.3 & 1640 & $7.5\times10^{-8}$ \\
225-25 & 22500 & 28.8 & 1130 & $7.4\times10^{-8}$ \\
200-25 & 20000 & 36.4 &  760 & $7.9\times10^{-8}$ \\
175-25 & 17500 & 47.6 &  570 & $1.4\times10^{-7}$ \\
150-25 & 15000 & 64.8 &  510 & $3.1\times10^{-7}$ \\
125-25 & 12500 & 93.3 &  120 & $1.7\times10^{-7}$ \\
100-25 & 10000 & 146  &  480 & $8.8\times10^{-9}$ \\
\hline
\multicolumn{5}{c}{$M=40\,{M}_{\odot}$, $\log(L/L_\odot)=5.66$,
$\Gamma=0.27$}\\
275-40 & 27500 & 29.9 & 1300 & $3.4\times10^{-7}$ \\
250-40 & 25000 & 36.1 & 1600 & $2.0\times10^{-7}$ \\
225-40 & 22500 & 44.6 & 1160 & $2.0\times10^{-7}$ \\
200-40 & 20000 & 56.4 &  700 & $2.6\times10^{-7}$ \\
175-40 & 17500 & 73.7 &  630 & $6.3\times10^{-7}$ \\
150-40 & 15000 & 100  &  110 & $1.5\times10^{-6}$ \\
125-40 & 12500 & 145  &   80 & $6.6\times10^{-7}$ \\
100-40 & 10000 & 226  &  410 & $2.9\times10^{-8}$ \\
\hline
\multicolumn{5}{c}{$M=60\,{M}_{\odot}$, $\log(L/L_\odot)=5.88$,
$\Gamma=0.30$}\\
275-60 & 27500 & 38.5 & 1240 & $6.1\times10^{-7}$ \\
250-60 & 25000 & 46.5 & 1850 & $3.2\times10^{-7}$ \\
225-60 & 22500 & 57.5 & 1000 & $3.4\times10^{-7}$ \\
200-60 & 20000 & 72.7 &  760 & $4.3\times10^{-7}$ \\
175-60 & 17500 & 95.0 &  820 & $1.2\times10^{-6}$ \\
150-60 & 15000 & 129  &  490 & $2.1\times10^{-6}$ \\
125-60 & 12500 & 186  &  110 & $1.3\times10^{-6}$ \\
100-60 & 10000 & 291  &  120 & $6.8\times10^{-8}$ \\
\hline
\end{tabular}
\end{table}

To understand the influence of small-scale inhomogeneities (clumping) on the
derived parameters, we calculated an additional set of models that account for
optically thin inhomogeneities \citep[for details see][]{irchuch}. The clumping
is parameterized by a clumping factor, which is introduced as
$\cc={\langle\rho^2\rangle}/{\langle\rho\rangle^2}$, and where the angle
brackets denote the average over volume. The value of $\cc=1$ corresponds to a
smooth flow. We adopted the empirical radial clumping stratification from
\citet{najradchuch} and \citet{bouhil}
\begin{equation}
\label{najc}
\cc(r)=C_1+(1-C_1) \, e^{-\frac{\vel(r)}{C_2}},
\end{equation}
where $C_1$ is the clumping factor and the velocity $C_2$ determines the
location of the onset of clumping. We adopted $C_1=10$ close to the value for
which the empirical H$\alpha$ mass-loss rates of O stars agree with observations
\citep{cmfkont} and $C_2=100\,\kms$, which is a typical value derived in
\citet{najradchuch}. In Eq.~\eqref{najc} we insert the fit of the velocity of
the unclumped wind ($\cc=1$) in an improved polynomial form of \citet{betyna} as follows: 
\begin{equation}
\label{vrfit}
\tilde \vel (r)=\sum_i \varv_i\zav{1-\gamma\frac{R_*}{r}}^i,
\end{equation}
where $\varv_i$ and $\gamma$ are parameters of the fit given in
Table~\ref{bvelech}. We note that this formula gives a more precise fit to the
radius dependence of velocity calculated by the solution of hydrodynamic equations
than the commonly used $\beta$-velocity law for any $\beta$.

\begin{table}[t]
\caption{Fit parameters (in Eq.~\eqref{vrfit}) of the velocity of unclumped
model and derived values of the terminal velocity $v_\infty$ and mass-loss rate
$\dot M$ for the model with clumping for $\cc=10$.}
\centering
\label{bvelech}
\begin{tabular}{crrrcrc}
\hline
\hline
Model & \multicolumn{4}{c}{Parameters of the velocity fit} &
\multicolumn{2}{c}{$\cc=10$} \\
 & $\varv_1$ & $\varv_2$ & $\varv_3$ & $\gamma$ & $\vinfty$ & $\dot M$ \\
\hline  
275-25 & 3525 & $-$1492 &      & 1.041 & 1950 & $1.0\times10^{-7}$ \\
250-25 & 4479 & $-$3038 &      & 1.051 & 1550 & $8.7\times10^{-8}$ \\
225-25 & 3676 & $-$3025 &      & 1.061 &  990 & $1.0\times10^{-7}$ \\
200-25 & 2077 & $-$1409 &      & 1.060 &  930 & $1.6\times10^{-7}$ \\
175-25 &  832 &  $-$139 &      & 1.066 &  810 & $3.7\times10^{-7}$ \\
150-25 &  765 &  $-$115 &      & 1.070 &  430 & $4.9\times10^{-7}$ \\
125-25 &  618 & $-$1408 & 1115 & 1.090 &  110 & $1.8\times10^{-7}$ \\
100-25 &  986 &  $-$488 &      & 1.132 &  510 & $9.1\times10^{-9}$ \\
\hline  
275-40 & 2504 & $-$1072 &      & 1.061 & 1280 & $3.9\times10^{-7}$ \\
250-40 & 3939 & $-$2389 &      & 1.064 & 1630 & $2.4\times10^{-7}$ \\
225-40 & 2908 & $-$1791 &      & 1.073 & 1160 & $3.1\times10^{-7}$ \\
200-40 & 1313 &  $-$522 &      & 1.076 & 1150 & $7.3\times10^{-7}$ \\
175-40 &  752 &     162 &      & 1.073 &  720 & $1.5\times10^{-6}$ \\
150-40 &  632 & $-$1390 & 1034 & 1.100 &  120 & $1.9\times10^{-6}$ \\
125-40 &  775 & $-$3375 & 5414 & 1.107 &   80 & $7.2\times10^{-7}$ \\ 
100-40 &  957 &  $-$488 &      & 1.166 &  510 & $3.1\times10^{-8}$  \\
\hline  
275-60 & 2527 & $-$1093 &      & 1.061 & 1170 & $7.2\times10^{-7}$ \\
250-60 & 4293 & $-$2377 &      & 1.067 & 1940 & $3.6\times10^{-7}$ \\
225-60 & 2812 & $-$1960 &      & 1.070 & 1180 & $6.0\times10^{-7}$ \\
200-60 & 1423 &  $-$514 &      & 1.072 & 1200 & $1.3\times10^{-6}$ \\
175-60 &  890 &      96 &      & 1.071 &  750 & $2.8\times10^{-6}$ \\
150-60 &  793 &  $-$152 &      & 1.068 &  510 & $2.7\times10^{-6}$ \\
125-60 &  780 & $-$2220 & 2183 & 1.090 &  110 & $1.4\times10^{-6}$ \\
100-60 &  728 & $-$1762 & 1452 & 1.145 &  100 & $7.0\times10^{-8}$ \\
\hline
\end{tabular}
\tablefoot{Unit of velocities $v_1$, $v_2$, $v_3$, and $v_\infty$ is \kms\
and the unit of mass-loss rate is \msr.}
\end{table}

\section{Calculated wind models}

\begin{figure}[t]
\begin{center}
\resizebox{\hsize}{!}{\includegraphics{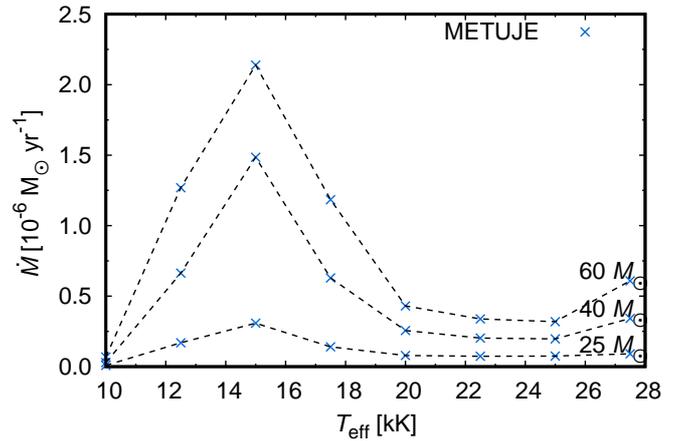}}
\end{center}  
\caption[]{Mass-loss rates $\dmdt$ without clumping predicted by our models
as a function of the stellar effective temperature \Teff.}
\label{dmdttep}
\end{figure}

\begin{figure}[t]
\centering
\resizebox{\hsize}{!}{\includegraphics{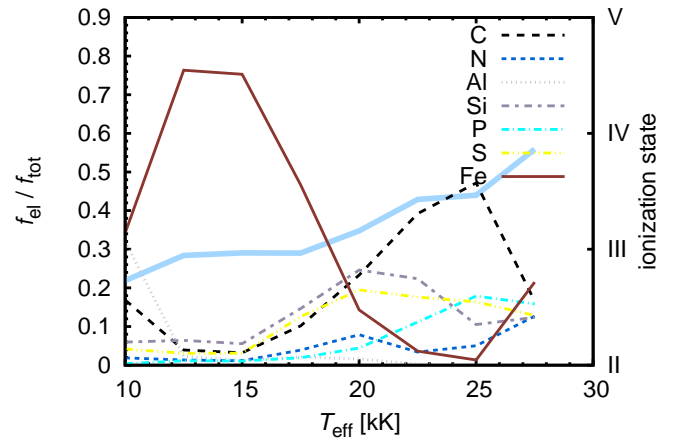}}
\caption{Relative contribution of individual elements to the line radiative
force at the wind critical point. Plotted as a function of the stellar effective
temperature for models with $M=40\,M_\odot$. The thick blue line plots the mean
ionization state that contributes to the line radiative force. This is defined
as $\sum f_iz_i/(\sum f_i)$, where $z_i$ is the ionization state of ion $i$ and
$f_i$ is its contribution to the line radiative force. This plot was derived in
a simplified way using the Sobolev approximation.}
\label{bvsil}
\end{figure}

Table~\ref{bvele} lists the input parameters of wind models and corresponding
wind parameters calculated without clumping. The wind mass-loss rate $\dmdt$
scales mostly with the stellar luminosity. At fixed luminosity, the mass-loss
rate slightly decreases with decreasing effective temperature down to about
$\Teff=22.5\,$kK (Fig.~\ref{dmdttep}, see also \citealt{vinbisja}). This trend
can be explained by a gradual shift of the emergent flux from the ultraviolet
region (where most of driving lines appear) to the visible region
\citep{vikola,btvit}. Around $\Teff=20\,$kK the mass-loss rate increases by a
factor of about 6 \citep{bista}. This rise of the mass-loss rate is usually
referred to as a bistability jump. It is caused by the iron recombination from
\ion{Fe}{iv} to \ion{Fe}{iii}, which accelerates wind more efficiently
\citep{vikolabis,vinbisja}. The increase of the mass-loss rate is not
instantaneous, but gradual in the interval of about 5\,kK.

\begin{table}[t]
\caption{Parameters of the fit of the mass-loss rate in Eq.~\eqref{dmdtob}.}
\label{fit}
\begin{tabular}{*{14}{c}}
\hline
\hline
$a$ & $b$ & $c$ & $T_1$ & $T_2$ & $\Delta T_1$ & $\Delta T_2$ \\
&&& \multicolumn{4}{c}{[kK]} \\
\hline
 $-$24.228 & 1.50 & 5.82 & 14.1 & 37.3 & 4.88 & 58.8 \\
\hline
\end{tabular}
\centering
\end{table}

The mass-loss rates of B supergiants and O stars from \citet{cmfkont} can be
simultaneously fitted via 
\begin{multline}
\label{dmdtob}
\log\zav{\frac{\dot M}{1\, \msr }}= a + b \log\zav{\frac{L}{10^6L_\odot}} \\
-a \log\szav{\exp\hzav{-\frac{\zav{T-T_1}^2}{\Delta T_1^2}}+
c \exp\hzav{-\frac{\zav{T-T_2}^2}{\Delta T_2^2}}}.
\end{multline}
The parameters of the fit are given in Table~\ref{fit}. The fit provides
an approximation for predicted mass-loss rates with a typical error of about 20 \%.

The temperature variations of the mass-loss rate can be also described using
Fig.~\ref{bvsil}, which plots the contribution of individual elements to the
line radiative force and the mean ionization degree contributing to that force.
For higher temperatures, $T_\text{eff}>20\,$kK, the wind is driven mostly by
lighter elements C, Si, P, and S. As the wind ionization decreases with
decreasing temperature for $T_\text{eff}<20\,$kK, iron recombines from
\ion{Fe}{iv} to \ion{Fe}{iii} and takes over most of the wind acceleration. This
results in a strong increase of the mass-loss rate explaining the
``bistability-jump'' behavior of hot star winds \citep{bista,vikolabis}.

Table~\ref{bvelech} gives the parameters of the fit of the wind velocity via
Eq.~\eqref{vrfit}. In most cases, the wind velocity is only described precisely enough by two
velocity parameters, $v_1$ and $v_2$. The parameter $\gamma$ is
always close to 1. This means that the size of nearly hydrostatic photosphere,
which is about $(\gamma-1)R_*$, is relatively small despite the supergiant
classification of all model stars. This can also be described in terms of the
ratio of the density scale height $H$ to the stellar radius, $H/R_\ast\sim
T_\text{eff}/(gR_\ast)$, which scales as $H/R_\ast\sim 1/T_\text{eff}$ for stars
at a constant luminosity. Therefore, this ratio varies roughly by a factor of
about 3 for studied stars and does not exceed 0.1, as can also be seen from
variation of parameter $\gamma-1$ with temperature in Table~\ref{bvelech}. This
differs from central stars of planetary nebulae, which also evolve at constant
luminosity, but which have significantly lower radii, consequently, for cool
central stars of planetary nebulae the atmospheric density scale height is
comparable with their radius \citep{btvit}.

Wind terminal velocity $\vinfty$ is predicted to be proportional to the escape
speed $\vesc$ \citep{cak}. For stars at fixed luminosity, the stellar radius
increases with decreasing effective temperature. Therefore, the escape speed and
wind terminal velocity become lower for cooler stars. However, the predicted
relation between $\vinfty$ and $\vesc$ is more complex. The wind terminal
velocity significantly decreases with temperature especially in the region
around $\Teff=20\,$kK \citep{vinbisja}. The decrease of the wind terminal
velocity in this region can be explained from the point of view of the line
strength distribution function \citep [see also Appendix~\ref{apalfa}] {pusle}.
With decreasing temperature, wind becomes progressively more accelerated by the
iron lines (corresponding to lower $\alpha$ in Fig.~\ref{alfa}). These lines are
weaker than the lines of lighter elements, many of which become optically thin in
the outer wind region and therefore do not accelerate the wind so efficiently to
high terminal velocities.

\begin{table}[t]
\caption{Parameters of the fit of the ratio of the terminal velocity to the
escape speed in Eq.~\eqref{vfit}.}
\label{vfittab}
\begin{tabular}{*{14}{c}}
\hline
\hline
$v_+$ & $v_-$ & $T_0$ [kK] & $\Delta T$ [kK]\\
\hline
3.0 & 0.7 & 20.7 & 3.3\\
\hline
\end{tabular}
\centering
\end{table}

On average, the ratio of wind terminal velocity and escape speed can be fitted
as
\begin{equation}
\label{vfit}
\frac{v_\infty}{v_\text{esc}}=
\frac{1}{2}\hzav{\zav{v_+-v_-}\frac{t}{1+|t|}+v_++v_-},\qquad
t=\frac{T_\text{eff}-T_0}{\Delta T},
\end{equation}
where the fit parameters $v_+$ and $v_-$ correspond to the limit of ratio at the
hot and cool end of the grid, $T_0$ corresponds to the jump temperature, and
$\Delta T$ to the width of the jump. The parameters derived from the fit of our
numerical results are given in Table~\ref{vfittab}.

The terminal velocity is around $100\,\kms$ for many models with
$\Teff\lesssim12.5\,$kK. This velocity could correspond to the slow
(subcritical) solutions of line-driven wind \citep{cak}. In most models, we are
able to discriminate between slow and fast solutions (see Fig.~4a in
\citealt{abbvln}) except at low effective temperatures, where the fast solution
approaches the slow solution. We were unable to find the faster solutions at low
temperatures, consequently, the real terminal velocities could be slightly
higher.

\begin{table}[t]
\caption{Calculated number of ionizing photons per unit of surface area
$\log \zav{Q/1\,\text{cm}^{-2}\,\text{s}^{-1}}$.}
\label{pocq}
\begin{tabular}{lrrrrrr}
\hline\hline
Model & \multicolumn{3}{c}{TLUSTY} & \multicolumn{3}{c}{METUJE}\\
& \multicolumn{1}{c}{\ion{H}{i}} & \multicolumn{1}{c}{\ion{He}{i}}
& \multicolumn{1}{c}{\ion{He}{ii}} & \multicolumn{1}{c}{\ion{H}{i}}
& \multicolumn{1}{c}{\ion{He}{i}} & \multicolumn{1}{c}{\ion{He}{ii}} \\
\hline
275-25 &  22.88 & 20.31 & 11.78 & 22.08 & 19.61 & 10.01 \\
250-25 &  22.28 & 19.31 & 10.21 & 21.58 & 18.81 & 8.88  \\
225-25 &  21.60 & 18.23 & 8.63  & 21.37 & 18.34 & 7.74  \\
200-25 &  20.70 & 16.87 & 6.57  & 21.11 & 17.94 & 6.39  \\
175-25 &  19.76 & 15.45 & 4.13  & 21.10 & 17.11 & 2.36  \\
150-25 &  18.86 & 14.04 & 0.84  & 20.93 & 16.36 & 0.51  \\
125-25 &  17.77 & 12.05 &       & 20.10 & 13.63 &  \\
100-25 &  15.70 &  8.78 &       & 18.01 &  9.81\\
\hline  
275-40 &  23.09 & 20.64 & 12.12 & 22.66 & 20.25 & 10.72 \\
250-40 &  22.55 & 19.69 & 10.62 & 21.59 & 18.74 & 8.60  \\
225-40 &  21.84 & 18.54 & 8.89  & 21.35 & 18.21 & 7.32  \\
200-40 &  20.96 & 17.20 & 6.91  & 21.22 & 17.82 & 5.42  \\
175-40 &  19.88 & 15.58 & 4.28  & 21.31 & 16.98 & 2.49  \\
150-40 &  18.94 & 14.13 & 0.73  & 20.78 & 13.70 & 1.24\\
125-40 &  17.88 & 12.18 &       & 20.06 & 13.99 & 0.74\\
100-40 &  15.92 &  8.99 &       & 18.06 &  9.85 \\
\hline  
275-60 &  23.13 & 20.73 & 12.21 & 22.73 & 20.35 & 10.84 \\
250-60 &  22.62 & 19.79 & 10.71 & 21.63 & 18.78 & 8.62  \\
225-60 &  21.91 & 18.64 & 8.95  & 21.37 & 18.18 & 7.30  \\
200-60 &  21.06 & 17.33 & 7.04  & 21.22 & 17.77 & 5.33  \\
175-60 &  19.93 & 15.64 & 4.34  & 21.29 & 16.56 & 1.08  \\
150-60 &  18.97 & 14.16 & 0.70  & 20.89 & 15.93 & 0.88  \\
125-60 &  17.90 & 12.21 &       & 20.09 & 13.24 \\
100-60 &  13.84 &  5.07 &       & 18.24 &  9.81 \\
\hline
\end{tabular}
\end{table}

\newcommand\vlo{0.495}
\begin{figure*}[tp]
\centering
\resizebox{\vlo\hsize}{!}{\includegraphics{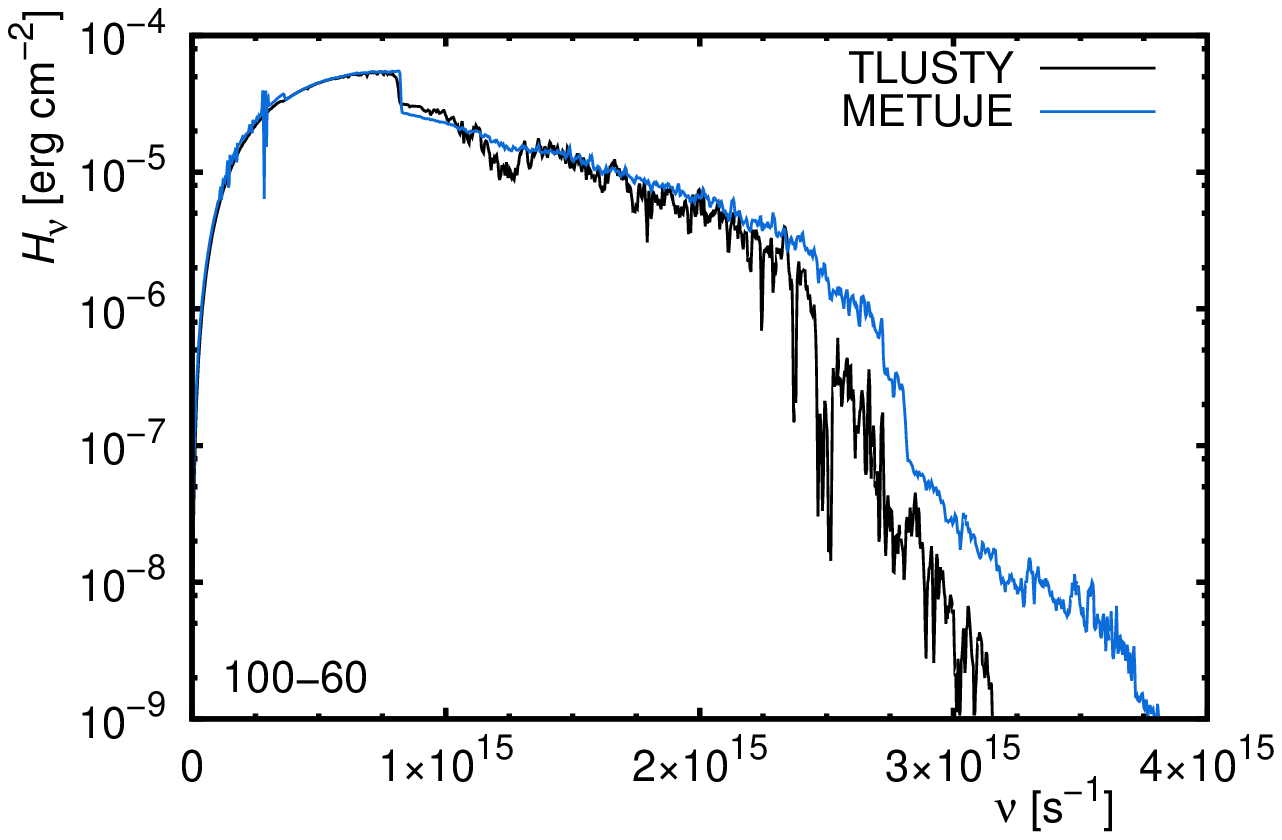}}
\resizebox{\vlo\hsize}{!}{\includegraphics{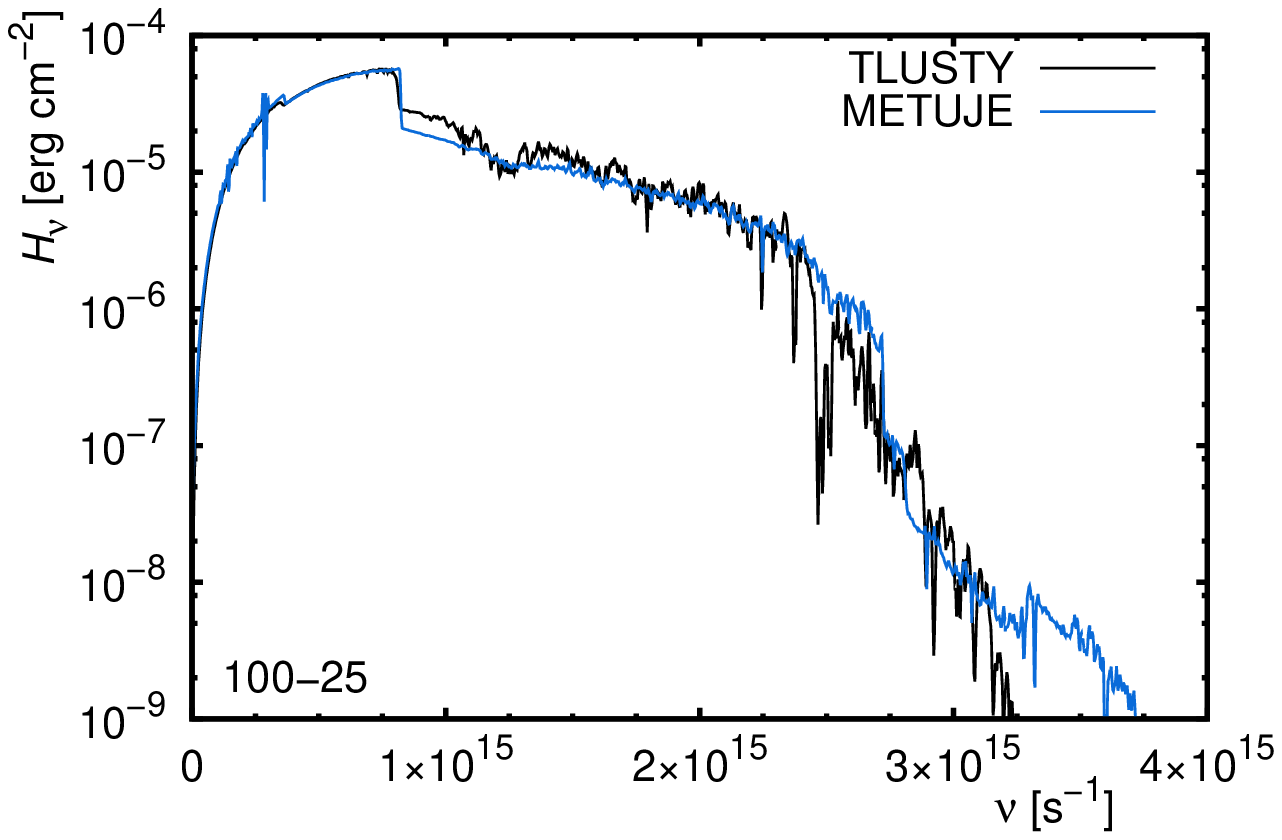}}
%\resizebox{0.32\hsize}{!}{\includegraphics{t12500g168y0851tok.eps}}
%\resizebox{0.23\hsize}{!}{\includegraphics{t12500g172y0851tok.eps}}
%\resizebox{0.32\hsize}{!}{\includegraphics{t12500g190y0851tok.eps}}
\resizebox{\vlo\hsize}{!}{\includegraphics{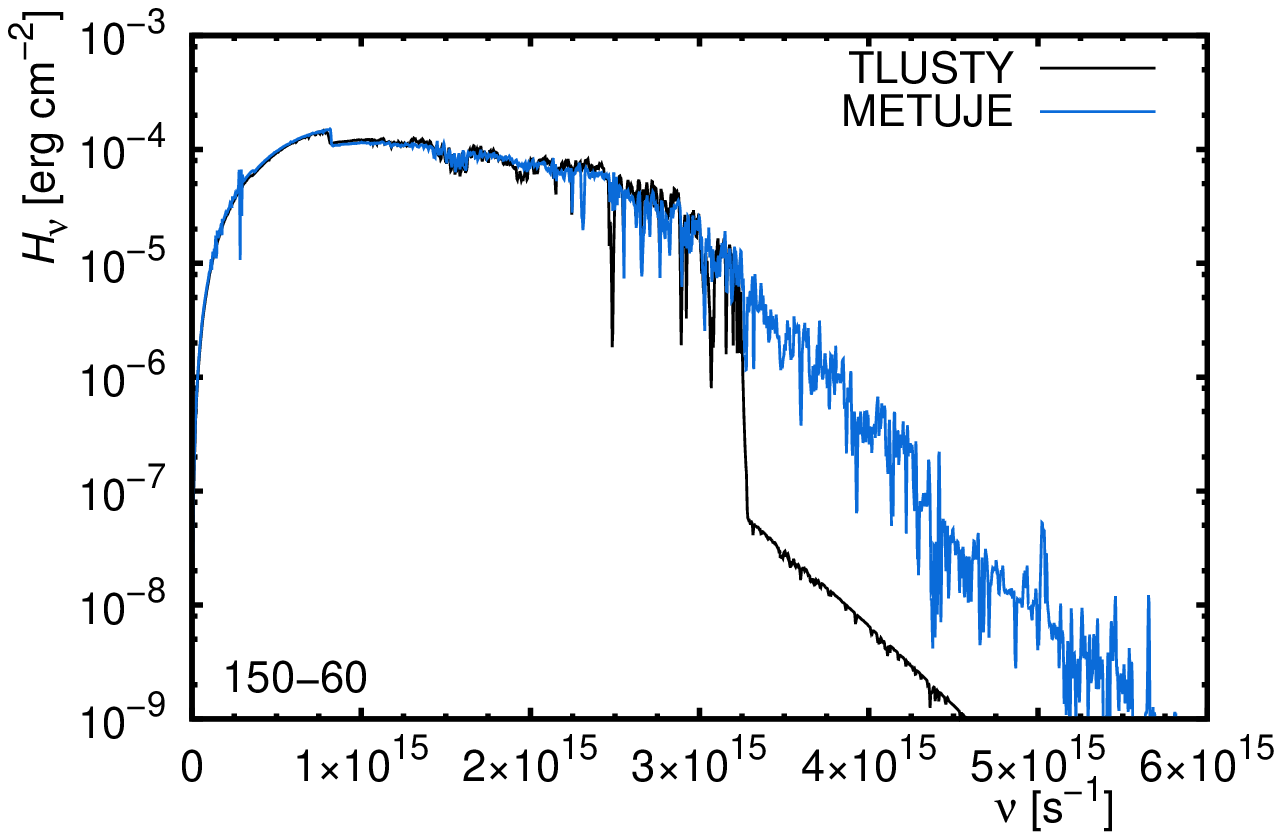}}
\resizebox{\vlo\hsize}{!}{\includegraphics{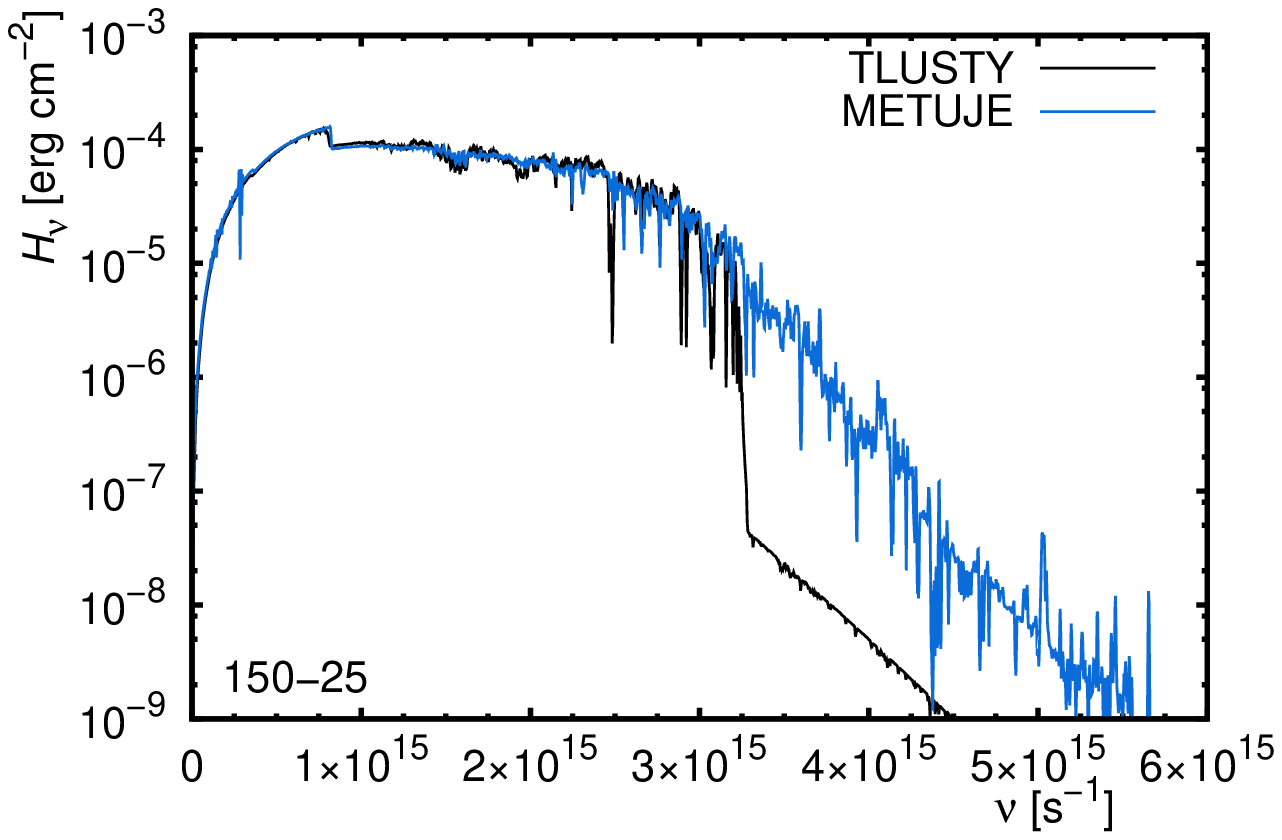}}
%\resizebox{0.32\hsize}{!}{\includegraphics{t17500g226y0851tok.eps}}
%\resizebox{0.23\hsize}{!}{\includegraphics{t17500g231y0851tok.eps}}
%\resizebox{0.32\hsize}{!}{\includegraphics{t17500g248y0851tok.eps}}
\resizebox{\vlo\hsize}{!}{\includegraphics{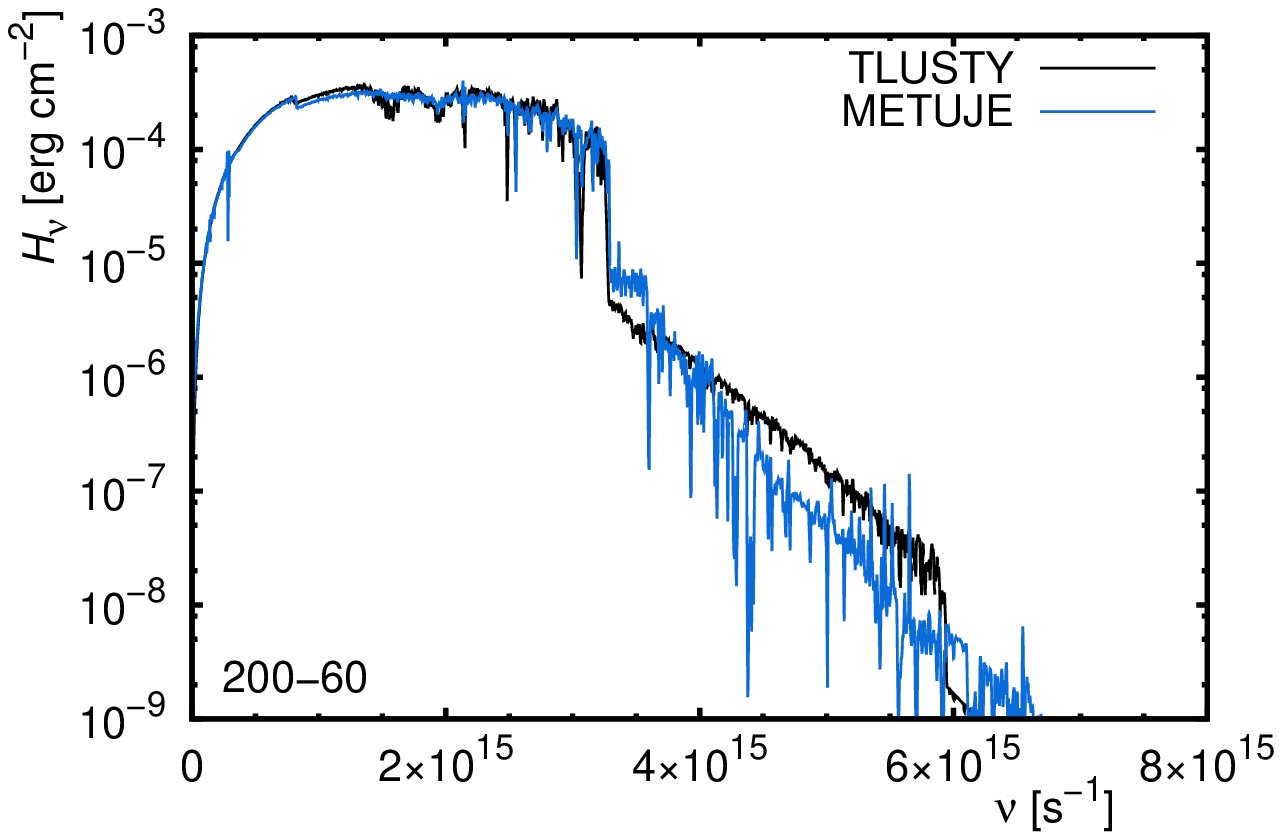}}
%\resizebox{0.23\hsize}{!}{\includegraphics{t20000g254y0851tok.eps}}
\resizebox{\vlo\hsize}{!}{\includegraphics{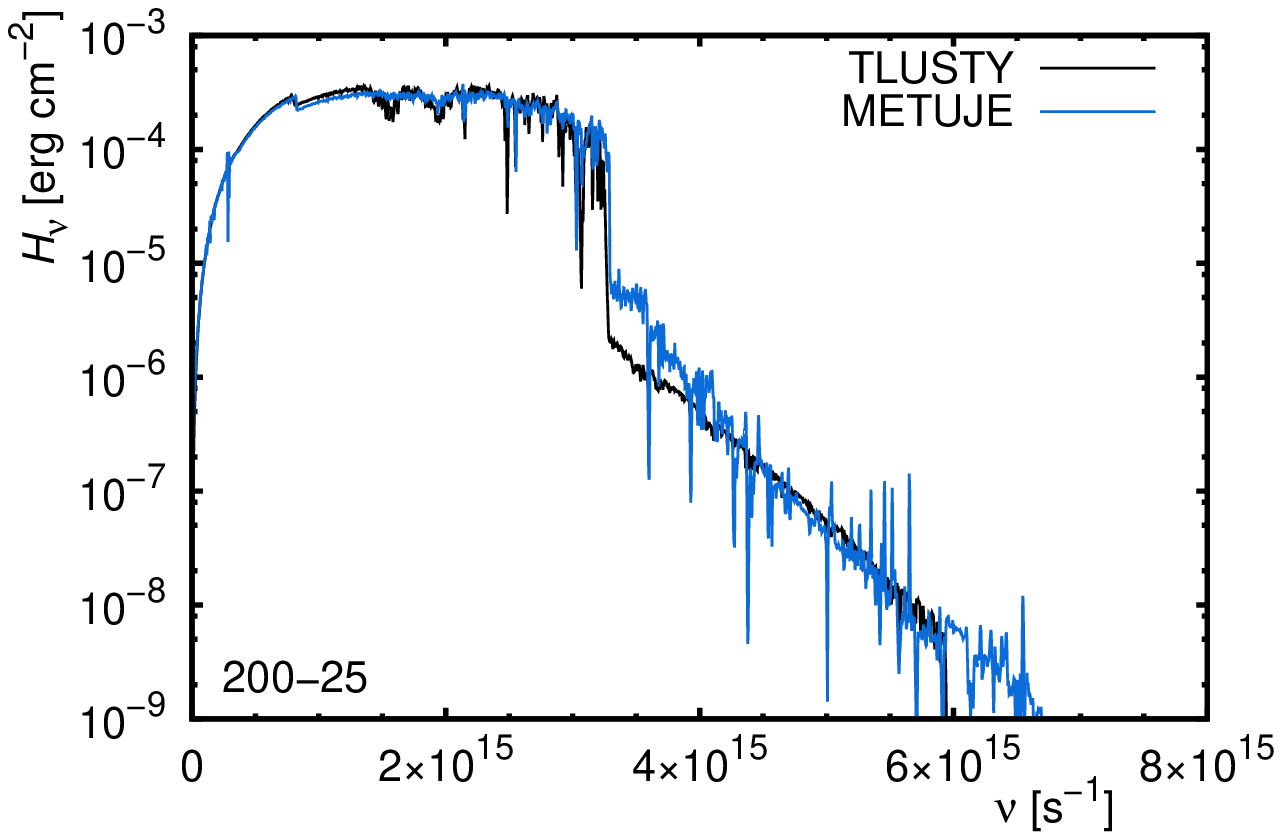}}
%\resizebox{0.32\hsize}{!}{\includegraphics{t22500g270y0851tok.eps}}
%\resizebox{0.23\hsize}{!}{\includegraphics{t22500g274y0851tok.eps}}
%\resizebox{0.32\hsize}{!}{\includegraphics{t22500g292y0851tok.eps}}
%\resizebox{0.32\hsize}{!}{\includegraphics{t25000g288y0851tok.eps}}
%\resizebox{0.23\hsize}{!}{\includegraphics{t25000g292y0851tok.eps}}
%\resizebox{0.32\hsize}{!}{\includegraphics{t25000g310y0851tok.eps}}
\resizebox{\vlo\hsize}{!}{\includegraphics{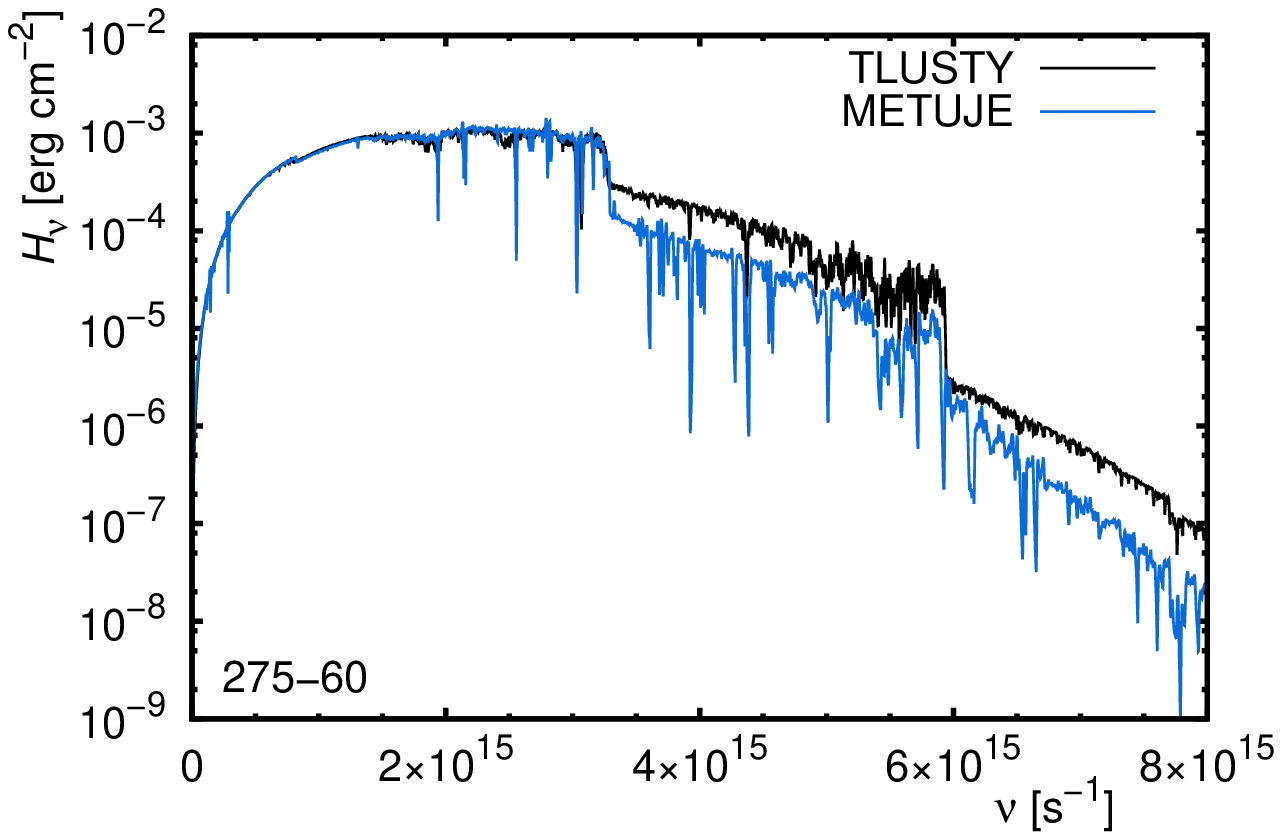}}
%\resizebox{0.23\hsize}{!}{\includegraphics{t27500g309y0851tok.eps}}
\resizebox{\vlo\hsize}{!}{\includegraphics{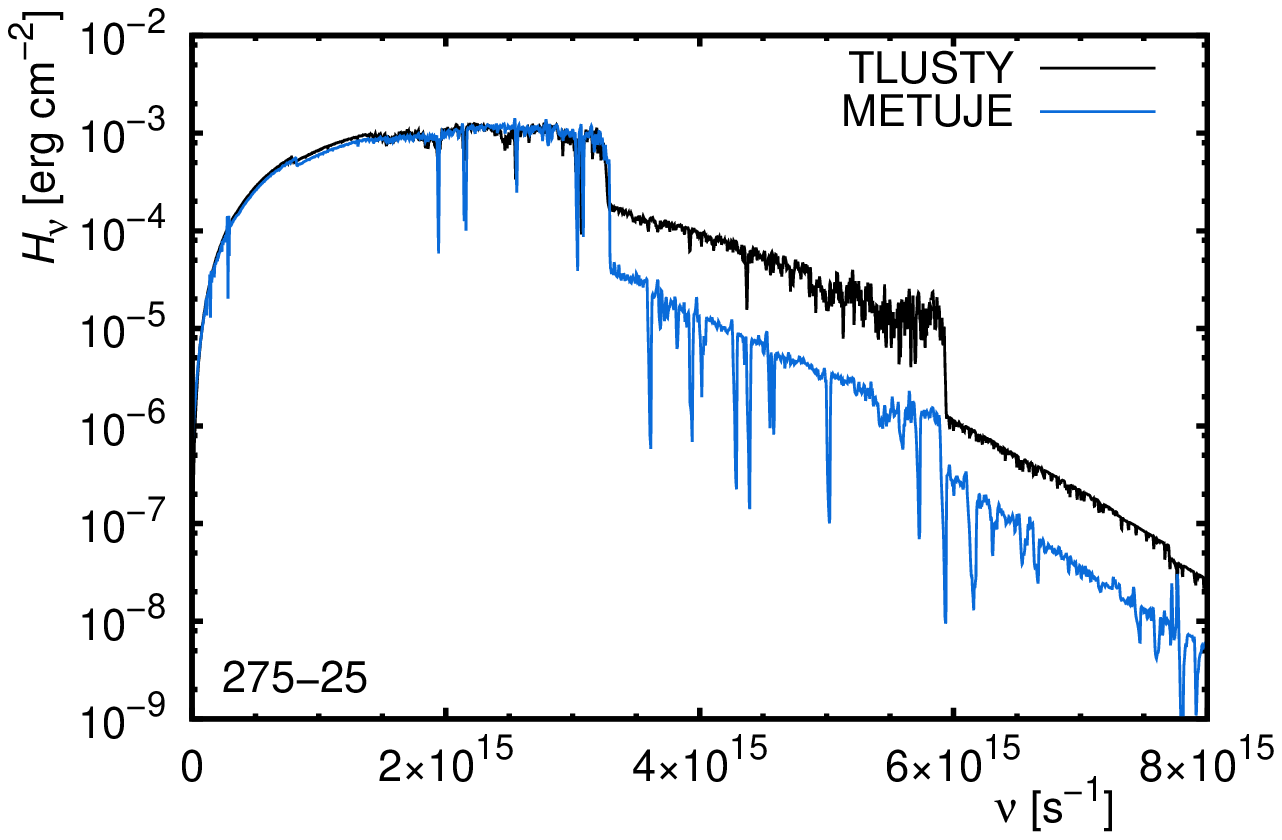}}
\caption{Comparison of the emergent fluxes from the plane-parallel TLUSTY (black
line) and global METUJE (blue line) models smoothed by a Gaussian filter. 
Stellar wind modifies the flux mostly above the Lyman jump.
Plotted for selected models from Table~\ref{bvele} (denoted in the plots).}
\label{bveletok}
\end{figure*}

In Fig.~\ref{bveletok} we compare emergent fluxes calculated using our global
models with fluxes derived from static plane-parallel photosphere models. These
fluxes nicely agree below the Lyman jump, because this part of continuum
originates in hydrostatic photosphere \citep{gableri}. The fluxes in this part
of spectra differ only in the atmospheres with very low gravity, where the
effect of sphericity becomes important \citep{kuhumi,grkud,gableri,kubii,scspn}.
The fluxes from hydrostatic and global models significantly differ for
frequencies higher than the Lyman jump, because at high frequencies the
continuum formation region moves into the wind.

With decreasing effective temperature, the ionizing flux in general decreases.
However, there are significant differences between far-ultraviolet fluxes from
the static plane-parallel models and global models with winds, which can to a
large extent be explained by the temperature bump appearing at low Rosseland
optical depths. This bump results from the Doppler shift of line centers from
their static positions leading to enhanced heating of the gas \citep{gableri}.
Further discussion of this interesting effect is given in \cite{fastpuls}. At the
highest effective temperatures considered 
%KrEd: here,
in our grid,
the far-ultraviolet continuum of
global models forms in the regions with lower temperature than in hydrostatic
models (below the temperature bump), therefore the flux in Lyman and \ion{He}{i}
continuum is lower in global models. Around $\Teff\approx20$\,kK the temperature
bump with nearly constant temperature becomes extended and as a result the jump
at $5.9\times10^{15}\,\text{s}^{-1}$ due to \ion{He}{i} disappears. For even
lower effective temperature the continuum forms in the region of the bump, which
is hotter than the corresponding region of the static plane-parallel atmosphere,
consequently, the far-ultraviolet flux also becomes higher.

The difference between emergent fluxes calculated from the plane-parallel and
global model atmospheres also appears in the number of ionizing photons emitted
per unit of surface area,
\begin{equation}
Q=4\pi\int_{\nu_0}^\infty \frac{H_\nu}{h\nu}\,\de\nu,
\end{equation}
given in Table~\ref{pocq}. In this equation, $H_\nu$ is the Eddington flux and the ionization
frequency $\nu_0$ stands for the ionization frequencies of \ion{H}{i},
\ion{He}{i}, and \ion{He}{ii}. While for hotter stars $\Teff\gtrsim40\,$kK the
plane-parallel models give a reasonable estimate of number of \ion{H}{i} and
\ion{He}{i} ionizing photons \citep{btvit}, for early B supergiants the
plane-parallel models overestimate the ionizing fluxes and for late B
supergiants underestimate ionizing fluxes with respect to global models. At the
hot end, the predicted number of ionizing photons corresponds to extrapolated
results for O supergiants from \citet{fastpuls} and from \citet{okali}. Our
ionizing fluxes slightly differ from those derived by \citet{smitioni} from
WM-BASIC models.

Wind parameters predicted from models calculated with clumping after
Eq.~\eqref{najc} are given in Table~\ref{bvelech}. Clumping increases the mass-loss
rate because it favors recombination and ions with lower charge drive wind more
efficiently \citep{muij,irchuch}. From the results in Table~\ref{bvelech}, the
wind mass-loss rates increase with clumping on average as $\cc^{0.2}$. The
effect of clumping is weaker than in O supergiants, in which the mass-loss
rate scales with clumping as $\dot M\sim\cc^{0.4}$ \citep{irchuch}. The lower
sensitivity of the mass-loss rate to clumping is connected with the fact that
radiative force in the range $22.5-27.5$\,kK is typically dominated by single
ionization state for most elements, therefore clumping does not significantly vary the
redistribution of the force among individual ionization states. On
the other hand, with clumping the increase of mass-loss rate resulting from iron
recombination appears at higher effective temperatures because the clumping
shifts the onset of the recombination from \ion{Fe}{iv} to \ion{Fe}{iii}.

We further tested the influence of $C_2$ parameter in Eq.~\eqref{najc} to
understand whether a weaker sensitivity to clumping is connected with the value
of this parameter, which determines the position of the onset of clumping. The
tests showed that this is not the case. For most models, a lower parameter
$C_2=30 \, \kms$ leads just to a slight increase of the wind mass-loss rate.

\section{Comparison with observations and other theoretical models}

\begin{figure}[t]
\begin{center}
\resizebox{\hsize}{!}{\includegraphics{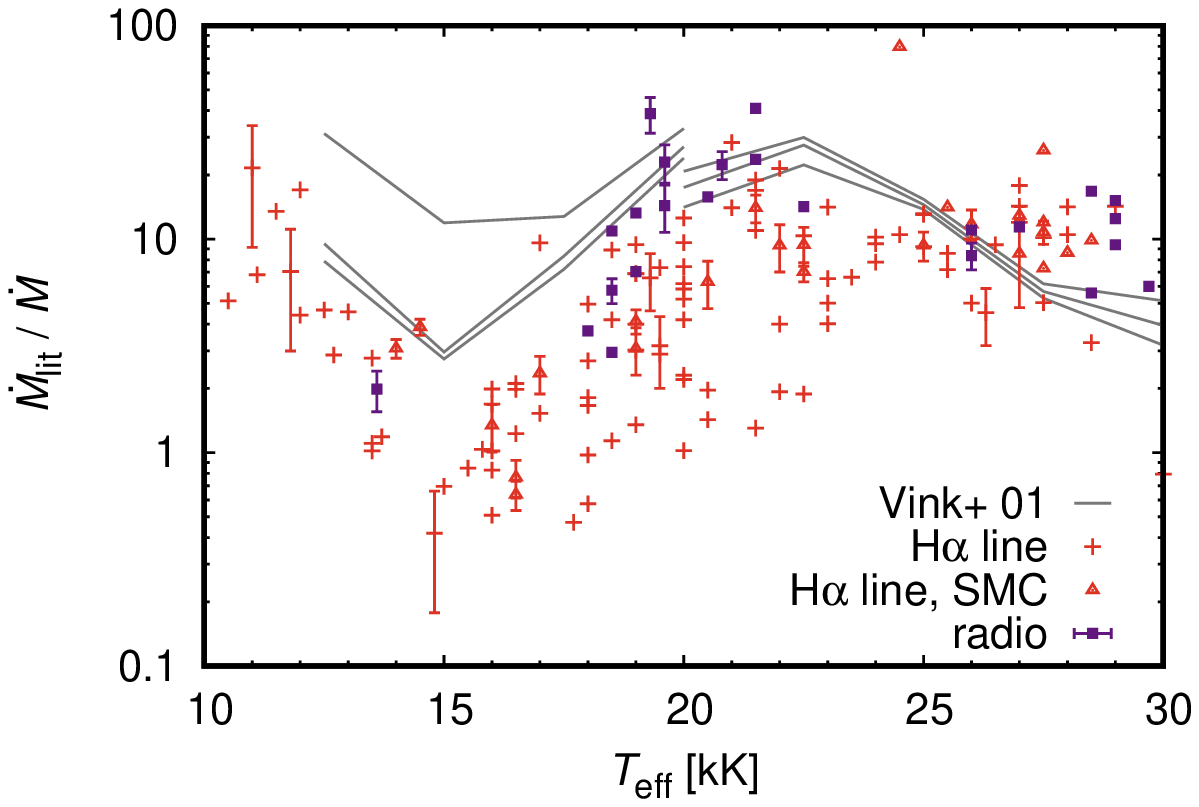}}
\end{center}  
\caption[]{Ratio of published mass-loss rates derived from observations using
analysis that neglects clumping ($\dmdt_\text{lit}$), and mass-loss rates
predicted using Eq.~\eqref{dmdtob}, plotted as a function of effective
temperature $\Teff$ for B supergiants. The observations include Balmer line
(mostly H$\alpha$) mass-loss rates of Galactic B supergiants \citep [red plus
symbols] {kudmomba,crow,lefever,marpulbvele,hauci}, H$\alpha$ mass-loss rates of
B supergiants from the Small Magellanic Cloud \citep [red triangles]
{bezchuch,trundlebsmc1,trundlebsmc2} scaled according to $\dot M\sim Z^{0.59}$
\citep{mcmfkont}, and mass-loss rates from radio data \citep[violet
squares]{scupamoc,benagek}. The ratios of the predictions of
\citet{vikolamet} to our predictions for three different luminosities are overplotted (solid
gray lines). Plotted with $v/v_\text{esc}=2$ for $\Teff\geq20\,$kK and with
$v/v_\text{esc}=1.3$ for $\Teff\leq20\,$kK.}
\label{dmdt_porov_crow} 
\end{figure}

\begin{figure}[t]
\begin{center}
\resizebox{\hsize}{!}{\includegraphics{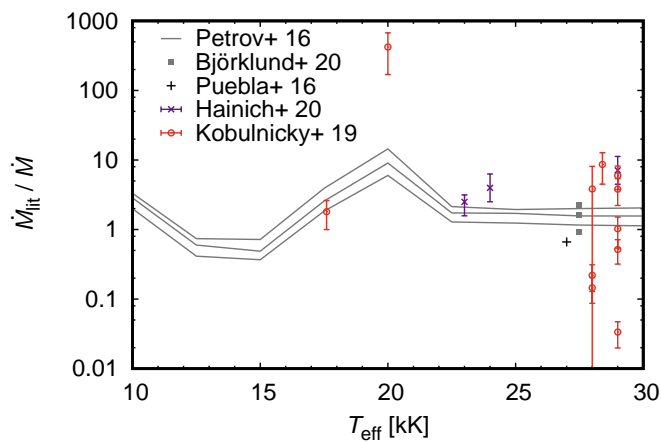}}
\end{center}  
\caption[]{Ratio of observational ($\dmdt_\text{lit}$) and predicted ($\dmdt$)
mass-loss rates as a function of effective temperature $\Teff$ for B
supergiants. Only observational mass-loss rates that are either
independent of clumping or were corrected for clumping in some way are plotted. The
observational results are based on X-ray line profiles \citep{vesnice}, UV
spectrum analysis with clumping \citep{haihmxb}, and wind bow shocks
\citep{kobul}. The gray lines denote predictions of \citet{petcmfgen} for different
luminosities and masses and the gray squares correspond to predictions of
\citet{bjorko} and $\Teff=27\,500$\,K.}
\label{dmdt_porov_pres} 
\end{figure}

The test of predicted wind mass-loss rates against the observations is not
straightforward owing to the effect of clumping on the observational wind
indicators. Small-scale wind inhomogeneities (clumping) are expected as a consequence of
strong radiatively driven wind instability \citep{ocr,felpulpal,felto}. Clumping
shifts the ionization equilibrium \citep{hamko,bourak,martclump,pulchuch} as a
result of stronger recombination in overdense regions. This affects the
classical wind mass-loss rate indicators such as H$\alpha$ line, ultraviolet wind
lines, and infrared and radio excess. The inferred values of wind mass-loss
rates spoiled by stronger recombination can be corrected for clumping by
multiplication by the factor of $1/\cc^{1/2}$. However, this simple scaling
breaks down in the case of clumps that are optically thick either in continuum
\citep{lidarikala} or in lines \citep{lidaarchiv,chuchcar,sund,clres1,clres2}.

Fig.~\ref{dmdt_porov_crow} compares predicted mass-loss rates with results
derived from Balmer lines and mass-loss rates from radio data. None of these
observational estimates account for clumping. For stars with
$T_\text{eff}>20\,$kK, the observational results are roughly by a factor of 9
higher than the predicted values. Assuming that the observational results should
be scaled down by a factor of $\sqrt\cc$ and that the predicted result increase
by a factor of $\cc^{0.2}$, the observational and predicted values would agree
for a higher value of $\cc$ than we used in our models, namely $\cc=24$. 

For stars with $\Teff<20\,$kK, the agreement between the observational and
predicted mass-loss rates is better (Fig.~\ref{dmdt_porov_crow}).  Although
there is a group of stars that show mass-loss rates significantly above the
theoretical predictions, the ratio of observational and theoretical mass-loss
rates decreases as a function of effective temperature and approaches 1 around
the mass-loss rate maximum at about $\Teff=15\,$kK. This implies that a lower
clumping factor of $\cc=5$ (on average) is required to bring observations and
theory to agreement. This conclusion agrees with the results derived from the
\mbox{CMFGEN} code \citep{pethalfa}, according to which the H$\alpha$ line is
only weakly sensitive to (optically thin) clumping for cool B supergiants.
Therefore, the H$\alpha$ mass-loss rates should be close to the theoretical
expectations. 

Besides determinations that are sensitive to clumping, there are methods that
yield mass-loss rates that are either independent of clumping or can be
corrected for clumping. Such rates are plotted in Fig.~\ref{dmdt_porov_pres}. We
plot the mass-loss rate (for $\epsilon$~Ori) derived from X-ray line profiles
\citep{vesnice}, which is independent of the effect of optically thin clumps,
but may be affected by optically thick inhomogeneities \citep{lidarikala}.
Fig.~\ref{dmdt_porov_pres} also includes mass-loss rates of B supergiant
components of high-mass X-ray binaries determined from UV line profiles
corrected for optically thin clumps \citep{haihmxb} and mass-loss rates of B
supergiants derived from stellar bow shocks \citep{kobul}. On average, these
observational estimates are by a factor of 1.4 higher than our predictions
(disregarding a star with $\Teff=20\,$kK). Therefore, the observational
mass-loss rates that are not affected by clumping agree much better with our
predictions, although significant discrepancies exist even here.

\begin{figure}
\includegraphics[width=0.5\textwidth]{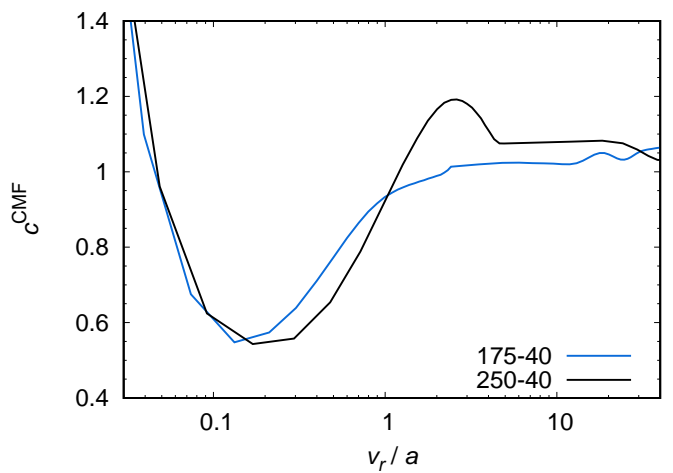}
\caption{Ratio of CMF line force and line force calculated using the Sobolev
approximation (denoted ad $c^\text{CMF}$)
as a function of velocity evaluated in terms of the depth dependent hydrogen
thermal speed $a=(2kT/m_\text{H})^{1/2}$.}
\label{ccmfb}
\end{figure}

Figures \ref{dmdt_porov_crow} and \ref{dmdt_porov_pres} also compare our results
with predictions of \citet{vikolamet}, \citet{petcmfgen}, and \citet{bjorko}.
From Fig.~\ref{dmdt_porov_crow} it follows that our predicted mass-loss rates
are significantly lower than predictions of \citet{vikolamet} by a factor of 13
and 8 for the models with $\Teff>20\,$kK and $\Teff<20\,$kK, respectively. This
difference can be attributed to the effect of line overlaps and to the Sobolev
approximation used by \citeauthor{vikolamet} to calculate the line force
\citep{cmfkont}. The latter is demonstrated in Fig.~\ref{ccmfb}, where we
plot the ratio of CMF and Sobolev line forces as a function of velocity. This
ratio is close to 1 for velocities on the order of a few thermal speeds, which
means that the Sobolev approximation is valid in that case. However, the ratio shows a
deep minimum around $v_r\approx0.1a$ ($a$ is a hydrogen thermal speed) owing to a
positive source function gradient near the photosphere \citep{cmf1}.

In Fig.~\ref{dmdt_porov_pres} we also overplot mass-loss
rates predicted using modified CMFGEN models \citep{petcmfgen}. These models
adopt similar assumptions as our code. Their results nicely agree with our
predictions with an exception of the regions around $\Teff\approx20\,$kK, where
\citet{petcmfgen} predict significantly larger mass-loss rates possibly due to
evolutionally earlier (i.e., for higher temperatures) onset of iron recombination
connected with the inclusion of clumping. \citet{bjorko} predict mass-loss
rates from the global version of the FASTWIND code with CMF line force. Their
model assumptions are very similar to ours, however their models were calculated
for O stars. Still, there is a near overlap of the grids around
$\Teff\approx28\,$kK, where the predictions reasonably agree.

\begin{figure}[t]
\begin{center}
\resizebox{\hsize}{!}{\includegraphics{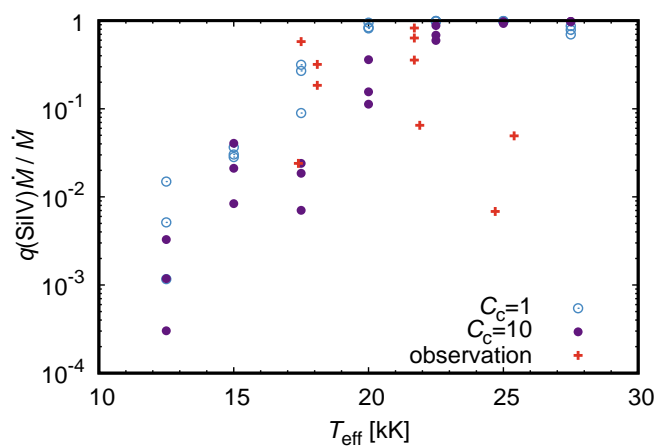}}
\end{center}  
\caption[]{Comparison of calculated and observed \citep{primanac} \ion{Si}{iv}
ionization fractions. The observed product of the \ion{Si}{iv} ionization
fraction and mass-loss rate $q(\ion{Si}{iv})\dot M$ was divided by the predicted
mass-loss rate to obtain the \ion{Si}{iv} ionization fraction. The calculated
ionization fractions are plotted with ($\cc=10$) and without ($\cc=1$) clumping
for $v=0.9\,v_\infty$.}
\label{ion_porov_prima}
\end{figure}

Clumps may directly affect the radiative transfer when they become optically
thick  (termed porosity and vorosity). This leads to an effectively gray
opacity of clumped media and an underestimation of mass-loss rates derived from
ultraviolet P~Cygni line profiles \citep{lidaarchiv,sund,clres2}. Therefore,
ratio of wind strengths of resonance doublet components can be used to test the
effect of clumping \citep{primacar}. If the ratio is close to the ratio of
oscillator strengths, then the porosity is negligible, while when it is equal to
1, then the porosity is significant.

\citet{primanac} applied this method to narrow absorption components, which
appear close to the blue edge of P~Cygni line profiles due to accumulation of
discrete absorption components. These authors concluded that these narrow absorption
components are not affected by porosity in B~supergiants and, consequently, are
suitable for the mass-loss rate determination and they estimated the product of
ionization fraction and mass-loss rate for studied stars, $q(\ion{Si}{iv})\dot
M$. In Fig.~\ref{ion_porov_prima} we plot these values for the stars for which
we derived luminosities either from the literature
\citep{kudmomba,crow,lefever,hauci} or determined using absolute magnitudes from
\citet{kalt} and bolometric corrections of \citet{crow}. The observational
values are plotted relative to the predicted mass-loss rate $\dmdt$ and
compared to predicted ionization fractions. The observational values of
$q(\ion{Si}{iv})\dot M$ are significantly lower than predictions for stars at
the hot side of the studied region, where \ion{Si}{iv} is a dominant ionization
stage. With decreasing effective temperature, silicon recombines from
\ion{Si}{iv} to \ion{Si}{iii}, improving the agreement between observations and
theory for lower effective temperatures. The recombination is stronger with
clumping, which explains the better agreement between the theory and observations in
Fig.~\ref{ion_porov_prima} with clumping in some cases. The huge disagreement at
the high effective temperatures is most likely caused by X-rays, which
significantly affect  the \ion{Si}{iv} ionization fraction \citep{nlteiii,lojza}.
Moreover, the radiative transfer in corotating interaction regions, where the
narrow absorption components are supposed to originate \citep{crowdac,lobl,duo},
is complex. Another problem is that only a small fraction of mass-loss is
carried by corotating interacting regions.

\begin{figure}[t]
\begin{center}
\resizebox{\hsize}{!}{\includegraphics{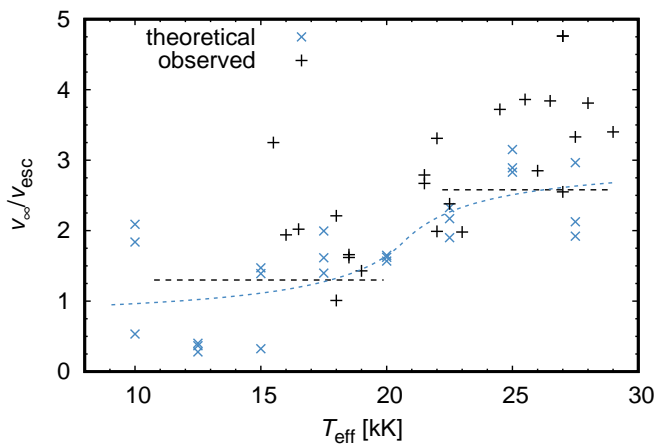}}
\end{center}  
\caption[]{Comparison of calculated and observed \citep{crow} ratios of the wind
terminal velocity and surface escape velocity. The dashed lines denote a mean
relationship of \citet{lsl} and the dotted blue line denotes mean predicted
ratio according to Eq.~\eqref{vfit}.}
\label{vnekvuni_crow} 
\end{figure}

The models are able to reasonably reproduce the wind terminal velocity. The wind
terminal velocity is proportional to the escape speed \citep{cak}, therefore the
ratio of these velocities does not show strong variations as a function of
effective temperature. In Fig.~\ref{vnekvuni_crow} we compare the predicted
ratio of the terminal velocity and escape speed with values derived from
observations.  The ratio decreases from about 3 to 1 as a result of
bistability jump, which agrees with observations \citep{lsl,crow} and
theoretical calculations \citep{vinbisja}.

\begin{figure}[t]
\begin{center}
\resizebox{\hsize}{!}{\includegraphics{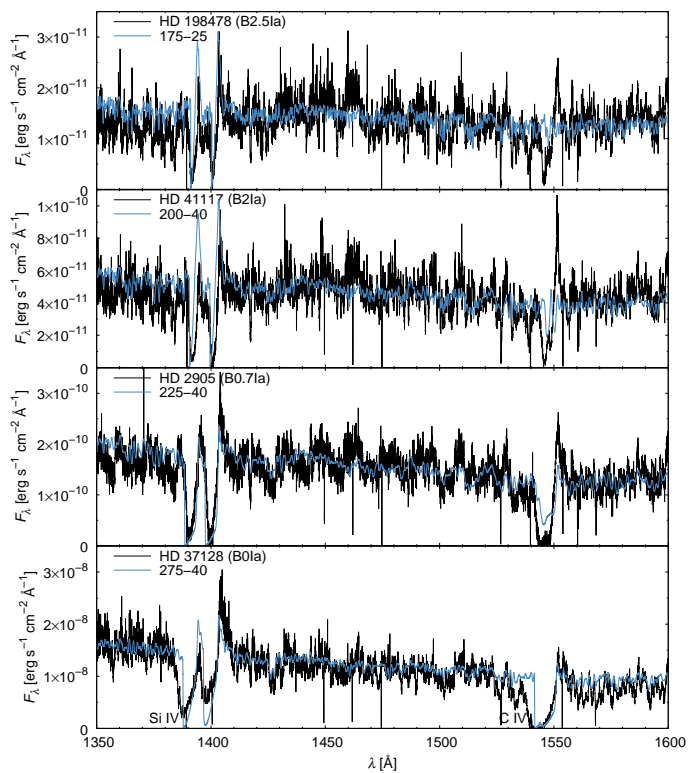}}
\end{center}  
\caption[]{Comparison of theoretical spectra from our grid and observed spectra
of selected B supergiants. The observed IUE spectra of HD~198478
(SWP~6335), HD~41117 (SWP~6338), HD~2905 (SWP~1814), and HD~37128 (SWP~8130) are plotted.}
\label{spektrabvele}
\end{figure}

To some extent, our models are also able predict the observed spectra (see
Fig.~\ref{spektrabvele}). The extent of the absorption part of the P~Cygni line
profiles is reasonably reproduced. This means that our models provide a reliable
estimate of the terminal velocities.  While the \ion{Si}{iv} doublet is also
typically reproduced well, the \ion{C}{iv} doublet is too weak for cooler stars.
We note that the comparison of individual observed spectra is done using our
limited grid of models and we do not aim for a detailed analysis of individual
objects.

\section{Discussion}

\subsection{Limitations of the models}

There are several limitations of our models that may affect the final results.
The most obvious limitation is the insufficiently accurate treatment of small-scale
wind structures, which are typically nicknamed as the effect of clumping.
Table~\ref{bvelech} demonstrates that optically thin inhomogeneities increase
the predicted mass-loss rates, while the optically thick inhomogeneities may
have the opposite effect, as described by \cite{muij}.

For our modeling, we only account for elements with the highest abundances,
while other elements are neglected \citep[see][for the list]{btvit}. This is
likely not a problem for lighter elements such as Cl or K, which do not
significantly contribute to the radiative force \citep{powrdyn}. Iron-peak
elements could possibly contribute to the radiative force more significantly as
a result of their higher number of lines, but these elements typically have at
least two orders of magnitude lower abundance than iron and nickel, which are
included in our models.

A potentially more significant problem may come from the number of lines that are
accounted for. Our line list is mostly based on observed lines, while
theoretical calculations give much higher number of lines \citep{carodej}. This
is not a problem for lighter elements, for which the line list was updated to
include even lines that are not observed. However, for iron the theoretical
line list includes a number of lines that is two orders of magnitude larger than our line
list, which may give rise to a larger error of the line force. To test this, we
compared line force calculated with shorter and longer iron line list and the
difference was typically just few percent. Slightly larger differences appeared
in the photospheric region, thus possibly affecting the structure of
the photosphere slightly, but not the mass-loss rate. Consequently, accounting for the
longer line list does not significantly affect our results.

\subsection{Bistability region}

Theoretical models agree in the prediction of a jump in wind mass-loss rates
\citep{bista,vikolabis,petcmfgen} and terminal velocities \citep{vinbisja}
around $\Teff\approx 21 $\,kK. This is called the bistability jump. Each of
these models applied different approximations to determine the quantities
characterizing the bistability jump, namely mass-loss rates and terminal
velocities, and yet derive comparable results. \citet[who discovered the
existence of the jump]{bista} used force multipliers to determine the radiation
force. \cite{vikolabis} combined several codes and succeeded in obtaining
significantly improved values of predicted mass-loss rates, and \cite{petcmfgen}
additionally used the code CMFGEN with CMF radiative transfer. However, the
observational support for the bistability jump is less clear. While the gradual
decrease of terminal velocities in the bistability region (a region of effective
temperatures where the bistability jump appears) is clearly supported by
observations (see Fig.~\ref{vnekvuni_crow}, \citealt{lsl,crow}), the presence of
a jump in mass-loss rates is unclear \citep{crow}. There is an indication of
a presence of a jump in radio wind efficiencies \citep{benagek}. The variability
of the mass-loss rates of luminous blue variables during their variation cycles
of S Doradus type can be described using varying mass-loss rates in the
bistability region \citep{vikolbv,grohilda}.

Our models are able to reproduce the observational behavior of mass-loss rates
provided that the clumping factor decreases from about 24 at the hot side of the
jump to about 5 at its cool side. Stronger clumping (higher clumping factor) at
the hot side of the bistability region can be connected with decrease of
macroturbulent velocity with decreasing effective temperature detected by
\cite{dufbvele} and \cite{marpulbvele}. At the cool side of the bistability
region, the difference between predicted and observational mass-loss rates
decreases with effective temperature. The
wind terminal velocity also decreases with decreasing effective temperature, possibly decreasing the amount of
clumping and as a result moderating the effect of clumping on observational
values \citep{trijist}.

Theoretical models predict that the H$\alpha$ line becomes optically thick for
cool B supergiants, leading to the appearance of the classical P~Cygni line profile
\citep{pethalfa}. Therefore, the line may become sensitive to optically thick
clumping. This could mean that observational H$\alpha$ mass-loss rate
determinations are underestimated at the cool side of the bistability jump
because the lines become optically thick there \citep{sund,clres2}.

\subsection{Evolutionary implications}

Massive stars lose a significant fraction of their mass during the evolutionary
phases corresponding to luminous B stars. For example, evolutionary models
(employing \citet{vikolamet} mass-loss rates) predict that a solar-metallicity
star with an initial mass of $60\,M_\odot$ loses about $25\,M_\odot$ during the
evolutionary stage of B supergiants \citep{gromek}. On the other hand, our
models were calculated for relatively low values of the Eddington parameter
$\Gamma\approx0.2-0.3$ (see Table~\ref{bvele}). When the star becomes a luminous
blue variable, it approaches the Eddington limit and its mass-loss rate
significantly increases \citep{gravigam,vinbisja}. To maintain a large total
amount of mass lost during stellar evolution, 
%KrEd this
ehnanced mass loss close to the Eddington limit
can possibly compensate
for lower values of mass-loss rates predicted in this work. 
Moreover, stars in a
proximity of the Eddington limit may have a strong porous optically thick
outflow \citep{shavedi,owosha} or may experience explosive mass-loss events
\citep{owoex}.

Stars with a lower initial mass ($\lesssim30\,M_\odot$) do not lose a
significant fraction of their mass as a result of line-driven winds
\citep{renzovit}. Consequently, the reduction of mass-loss rates mentioned in this work
might not be significant for the evolution of these stars. On the other hand,
the compactness of the core of pre-supernova models is sensitive to mass-loss
during the hot evolutionary phase \citep{renzovit}. Compactness is one of
the parameters that determines an outcome of a supernova explosion
\citep{ocot,ugliano}.

Stars do not only lose mass via their winds, but also angular momentum.
\citet{vibrobra} proposed that low observed rotational velocities of cool B
supergiants could be caused by stronger angular momentum loss in the vicinity of
the bistability jump. \citet{kostel} tested several experimental wind
prescriptions and concluded that low rotational velocities of cool B supergiants
could be reproduced even with weaker mass-loss rates, but with a bistability
jump in mass-loss rates.

In the early evolutionary phases the wind velocity slows down with decreasing
effective temperature and while the star passes the bistability region, the
mass-loss rate increases. In later evolutionary phases, for example during the
Wolf-Rayet phase, the wind velocity increases again. Consequently, the faster
wind in later evolutionary phases may collide with the wind from earlier
evolutionary phases and create circumstellar structures similar to those found
in planetary nebulae \citep{kwok}.

\subsection{Domains of radiatively driven mass loss}

\begin{figure}
\begin{center}
\resizebox{\hsize}{!}{\includegraphics{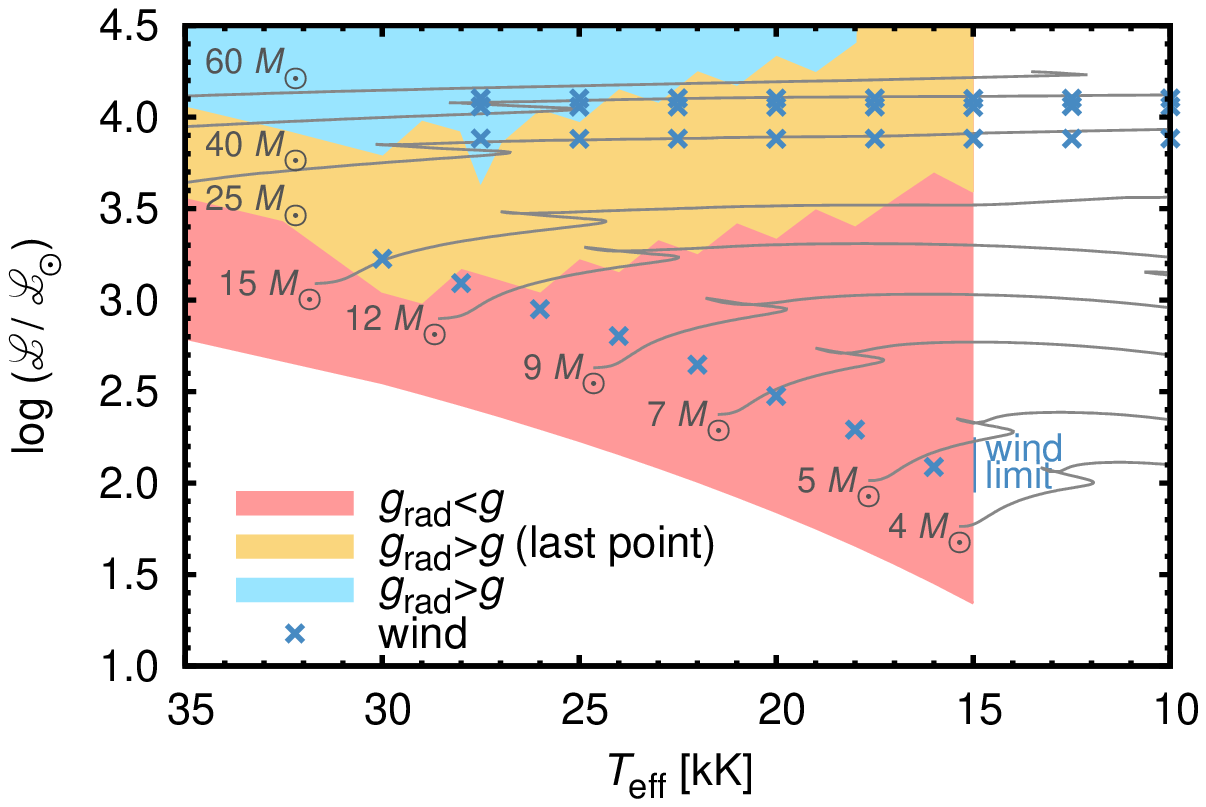}}
\end{center}  
\caption[]{Domains of radiatively driven mass loss in the spectroscopic
Hertzsprung-Russell diagram. The domains were derived from the balance of
gravity and radiative force using the OSTAR2002 and BSTAR2006 static model
atmosphere grids \citep[][respectively]{ostar2003,bstar2006}. In this diagram,
$\text{\fontfamily{frc}\itshape \fontsize{6pt}{8pt}\selectfont
L}=T_\text{eff}^4/g$. The red domain denotes parameters where the hydrostatic
photosphere is possible, the filled yellow region indicates the parameters where the
hydrostatic photosphere is possible and the radiative force overcomes gravity
only in the last point of model photosphere, and the blue area corresponds to
the parameters where the hydrostatic photosphere is impossible. The white area
in the right and bottom parts of the figure reflects no corresponding models exist in the static model atmosphere grids used. The blue
crosses denote locations of wind models from Table~\ref{bvele} and
\citet{metuje}. The solar-metallicity evolutionary tracks of
\citet{sylsit} are overplotted.}
\label{speltef}
\end{figure}

\citet{abbobla} introduces two limits (static and wind limit, see his Fig.~1)
and discriminates between three zones in the Hertzsprung-Russell diagram. Above
the static limit, fully hydrostatic photospheres are not possible, the radiative
force overcomes gravity at some point, and the wind is self-initiated. Below the
static limit and above the wind limit, hydrostatic photospheres are possible,
therefore the wind is not self-initiated, but it can be sustained. Therefore,
two types of solutions exist in this region. Below the wind limit, only
hydrostatic photospheres are possible and the chemically homogeneous wind does
not exist.

To plot these wind domains in Fig.~\ref{speltef}, we used the spectroscopic
Hertzsprung-Russell diagram \citep{lkdiagram,revoldiagram} and overplotted the
results of models from OSTAR2002 and BSTAR2006 static model atmosphere grids
\citep{ostar2003,bstar2006}. For high luminosities, static atmospheres are not
possible, and the outer atmospheric levels expand triggering the wind. This case
corresponds to the atmospheres of B supergiants. However, for many B supergiants
the radiative force $g_\text{rad}$ exceeds the magnitude of gravity $g$ only in
single outermost layer of the model atmosphere. This layer may be the subject of
numerical inconsistencies, consequently, many B supergiants may lie
close to the static limit.

\subsection{X-ray emission}

\begin{figure}
\begin{center}
\resizebox{\hsize}{!}{\includegraphics{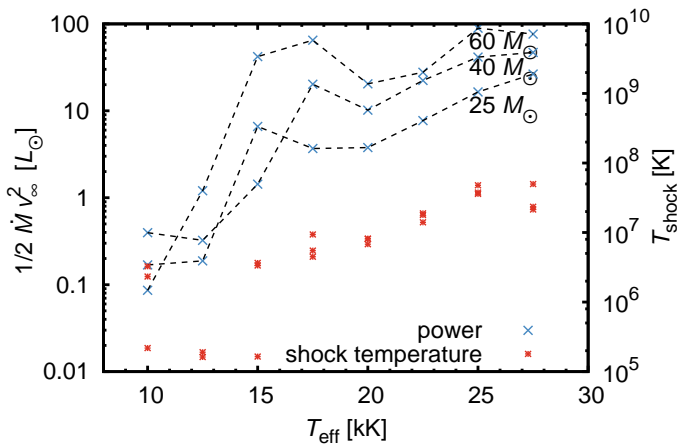}}
\end{center}
\caption[]{Wind power (left axis) and shock temperature (right axis)
as functions of stellar effective temperature.}
\label{vvykon}
\end{figure}

Hot stars emit X-rays, which in most cases originate in shocks thought to be
caused by line driven wind instability \citep{ocr,felpulpal}. In B supergiants,
the \mbox{X-ray} emission typically appears only in the earliest spectral types
up to about B3 \citep{rosat,naze}. This roughly corresponds to the effective
temperature of about 15\,kK \citep{marpulbvele}.

The wind X-ray emission is powered by the wind kinetic energy. In
Fig.~\ref{vvykon} we plot the wind power $\dot Mv_\infty^2/2$ as a function of
the effective temperature. The wind power does not significantly vary in hotter
B supergiants even in the region of recombination of \ion{Fe}{iv} to
\ion{Fe}{iii}, where the increase of the mass-loss rate is compensated by the
decrease of the wind terminal velocity. On the other hand, mostly as a result of
a decrease of the wind terminal velocity, the wind power significantly decreases
between 17.5 -- 12.5~kK, exactly in the region where the X-ray emission
disappears.

The temperature of the post-shock gas is proportional to the square of the
terminal velocity. As a result, with decreasing effective temperature the
post-shock gas temperature significantly decreases (Fig.~\ref{vvykon}) from
about 5\,MK for $\Teff=17.5\,$kK to about 200\,kK for $\Teff=12.5\,$kK (using
velocities from Table~\ref{bvele} and Eq.~(1) of \citealt{owomix}). Therefore,
the X-ray emission becomes negligible at such low temperatures
\citep[e.g.,][]{rs,florentani}.

Depending on the dominant cooling mechanism, shocks may be either radiative or
adiabatic. As a result of the decrease of the wind density with radius, shocks
are typically radiative close to the star and adiabatic in the outer wind parts.
Because of their large stellar radii, from Eq.~(24) of \cite{owomix} it follows that
the transition from radiative to adiabatic shocks appears in B supergiants
relatively close to the star at a typical distance of few stellar radii.

\subsection{Influence of the magnetic field}

Although the Zeeman splitting does not directly affect the wind mass-loss
rate for typical magnetic fields detected in OB stars \citep{malabla}, the
magnetic field still affects the mass-loss rate as a result of wind channelling
along the field lines \citep{udo}. The effect of the magnetic field scales with
the ratio of the magnetic field energy density to the wind kinetic energy
density, which at large distances from the star simplifies to the wind magnetic
confinement parameter of \citet{udo}
\begin{equation}
\label{eta}
\eta_\ast=\frac{B_\text{eq}^2R_\ast^2}{\dot Mv_\infty},
\end{equation}
where $ B_\text{eq}$ is the equatorial field strength. For weak confinement,
$\eta_\ast\lesssim1$, the magnetic field opens up and the wind flows radially.
On the contrary, for strong confinement, $\eta_\ast\gtrsim1$, the wind is
trapped by the magnetic field.

In rewriting Eq.~\eqref{eta} in terms of typical parameters of B supergiants, we
derive
\begin{equation}
\label{peta}
\eta_\ast=770\zav{\frac{B_\text{eq}}{100\,\text{G}}}^2
\zav{\frac{ R_\ast }{100\,R_\odot}}^2
\zav{\frac{\dot M}{10^{-7}\,M_\odot\,\text{yr}^{-1}}}^{-1}
\zav{\frac{v_\infty}{10^3\,\kms}}^{-1}.
\end{equation}
The field strength in typical magnetic O stars is on the order of 1\,kG
\citep{donhd191612,wadngc}. Assuming flux conservation, which is approximately
valid in more massive magnetic chemically peculiar stars \citep{otevrenky}, the
magnetic field strength in B supergiants is on the order of 10\,G. This is a
typical strength of magnetic fields found in evolved BA stars
\citep{fosbobvel,coravel,nemrazvel}. From Eq.~\eqref{peta}, it follows that such
fields are able to strongly confine the stellar wind.

The angular momentum loss from magnetized wind leads to rotational braking. In
principle, this could provide additional angular momentum loss required by
evolutionary models with low mass-loss rates to reproduce observed rotational
velocities \citep{kostel}. With an angular momentum constant $k$ on the order of
0.01 as derived from MESA evolutionary code \citep{mesa1,mesa2} for a B
supergiant stage of the star with initial mass $40\,M_\odot$, the magnetic
braking timescale is on the order of 1\,Myr using Eq.~(25) of \citet{brzdud} and
typical wind parameters found in this work. This is order of magnitude longer than the
duration of B supergiant stage \citep{gromek}. From this it follows that the
magnetic braking does not likely significantly affect the rotational speed of
B supergiants.

\subsection{Pure absorption models}

Analyzing the topology of wind solutions around the sonic point, \citet{havran}
concluded that pure absorption models have nodal topology instead of the saddle-type topology, which would correspond to solar wind. This leads to a strong
instability of time-dependent pure absorption wind models \citep{ocr} and may
prevent us from calculating time independent models. However, the line emission
may restore the saddle-type topology and provide stability for time independent
models \citep{sunowroz}.

To test this, we recalculated the wind model 275-60 neglecting line emission in CMF
radiative transfer equation and starting with  a converged model corresponding to a
relatively low mass-loss rate. It turned out that we were unable to
converge the pure absorption model. Therefore, the inclusion of the line
emission (scattering) is crucial to obtain well-converged wind models. We
further tested the solution topology around the sonic point, thereby concluding that the
wind has a saddle-type topology in the case with emission, but a nodal topology
without emission (see Appendix~\ref{sedlo}).

\section{Conclusions}

We provide global line-driven wind models for B supergiants at the metallicity
corresponding to our Galaxy. The structure of the flow is determined
consistently from the nearly hydrostatic photosphere to the wind expanding with
supersonic speed. The winds are driven mostly by C, Si, and S for hotter B
supergiants, while iron dominates wind driving in cooler supergiants. We
determined basic wind parameters, that is, the mass-loss rates and terminal
velocities.

We generalized our formula for fast estimates of mass-loss rate predictions of O
stars to B supergiants as well (Eq.~\ref{dmdtob}). The mass-loss rate depends mostly on
the stellar luminosity. The dependence of the mass-loss rate on the effective
temperature is nonmonotonic. The mass-loss rates decrease with temperature down
to a minimum at about 22.5\,kK. For cooler stars the mass-loss rates gradually
increase by a factor of about 6 in the region of bistability. The increase is
caused by recombination of iron from \ion{Fe}{iv} to \ion{Fe}{iii}. The
mass-loss rates reach maximum at about 15\,kK, where they start to decrease with
temperature again.

The wind terminal velocity is proportional to the escape speed for a given
effective temperature, as can be seen from our fitting formula Eq.~\eqref{vfit}.
The ratio of the terminal velocity to the escape speed decreases from about 2 --
3 at the hot side of the region of iron recombination to about 0.5 -- 1.5 at the
cool side. This nicely agrees with observations.

The comparison with mass-loss rates derived from observations is more
problematic as a result of the effect of clumping on the observational
indicators. The mass-loss rates that are uncorrected for clumping are by a
factor of 3 higher than the predictions on the cool side of the bistability,
while they are by a factor of 9 higher than the predictions at the hot side
of the bistability. These differences can be alleviated by the influence of
clumping on observational determinations and on the predictions. Hence clumping
that decreases with temperature can resolve the discrepancy between mass-loss
rate estimates affected by clumping and theoretical predictions \citep{trijist}.
On the other hand, the mass-loss rate estimates that are not sensitive to
clumping reasonably agree with our predictions. Mass-loss rates derived from
X-ray line profiles, UV spectroscopy that accounts from clumping, and stellar
bow shocks are on average just by a factor of 1.4 higher than our predictions.

The predicted mass-loss rates are by a factor of about 10 lower than values
being used in current evolutionary models. Together with previous studies that
predict a similar reduction of line-driven mass-loss rates
\citep{cmfkont,sundyn}, this result calls for a re-evaluation of the role of winds
in evolution of massive stars.

\begin{acknowledgements}
This work was supported by grant GA \v{C}R 18-05665S. Computational resources
were supplied by the project "e-Infrastruktura CZ" (e\nobreakdash-INFRA
LM2018140) provided within the program Projects of Large Research, Development
and Innovations Infrastructures. The Astronomical Institute Ond\v{r}ejov is
supported by a project RVO:67985815 of the Academy of Sciences of the Czech
Republic.
\end{acknowledgements}

%%%%%%%%%%%%%%%%%%%%%%%%%%%%%%%%%%%%%%%%%%%%%%%%%%%%%%%%%%%%%%%%%%%%%%%%

\bibliographystyle{aa}
\bibliography{papers}

\appendix

\section{Solution topology around the sonic point}
\label{sedlo}

As shown by \citet{holzer}, the solution topology can be determined from the
analysis of the momentum equation
\begin{equation}
\label{jestrebi}
vv'=\frac{a^2}{v}v'-\frac{GM}{r^2}+ g_\text{rad}+\frac{2a^2}{r}
\end{equation}
around the sonic point. In this equation, we neglected the radial variations of the
isothermal sound speed $a$, $g_\text{rad}$ denotes the radiative force, and
$v'=\frac{\de v}{\de r}$. The momentum equation Eq.~\eqref{jestrebi} yields for
the velocity gradient
\begin{equation}
v'=\frac{1}{v}\frac{g_\text{rad}-\dfrac{GM}{r^2}}{1-\dfrac{a^2}{v^2}},
\end{equation}
where we dropped the last term on the right-hand side of Eq.~\eqref{jestrebi}, which
is negligible for line-driven winds. At the sonic point, the velocity derivative
is derived using the L'Hospital's rule, which gives 
\begin{equation}
\label{jestrebilopital}
v'^2-\frac{1}{2}\pd{g_\text{rad}}{v}v'-\frac{1}{2}
\zav{\pd{g_\text{rad}}{r}+\frac{2GM}{r^3}}=0.
\end{equation}
We assumed that $g_\text{rad}$ is a function of radius and velocity only.
Close to the sonic point the derivative of the radiative force dominates in
the last bracket in Eq.~\eqref{jestrebilopital}, hence 
Eq.~\eqref{jestrebilopital}
has positive and negative roots implying a saddle-type topology if
\begin{equation}
\label{furtjenejstrebi}
\pd{g_\text{rad}}{r}>0,
\end{equation}
as was derived by \citet{havran}.

It is difficult to derive ${\partial g_\text{rad}}/{\partial r}$ directly from
numerical models. However, the models give $\de g_\text{rad}/\de r$ from the
radial dependence of the radiative force, which can be converted to the partial
derivative using
\begin{equation}
\pd{g_\text{rad}}{r}=\frac{\de g_\text{rad}}{\de r}-\pd{g_\text{rad}}{v}v'.
\end{equation}
The derivative ${\partial g_\text{rad}}/{\partial v}$ can be computed
numerically from two runs with slightly different velocity. We tested the
solution topology in the wind models with line emission using
Eq.~\eqref{furtjenejstrebi}. The partial derivative in
Eq.~\eqref{furtjenejstrebi} is positive implying saddle topology
around the sonic point,
as already shown from analytic calculations by \citet{sunowroz}.
We note that this holds even if the critical point of our wind models appears
downstream at the point where the wind velocity is equal to the speed of
radiative-accoustic waves \citep{abbvln,thofe}.

\section{Line strength distribution functions}
\label{apalfa}

The contribution of lines with different strengths to the radiative force can be
described using the line strength distribution function \citep{cak,pusle}. The
line strength distribution function can be conveniently parameterized using $k$,
$\alpha$, and $\delta$ parameters, which enable the simplification of the
calculation of the line force. \citet{pusle} showed that the $\alpha$ parameter
corresponds to the ratio of radiative force due to optically thick lines to the
total line force.

\begin{figure}[t]
\includegraphics[width=0.5\textwidth]{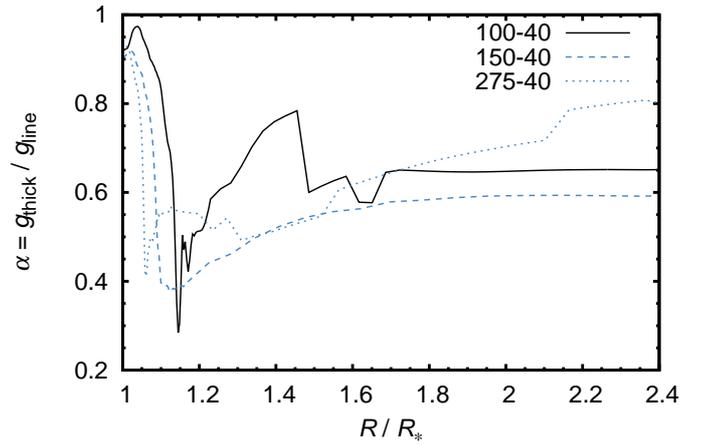}
\caption{Ratio of the radiative force due to the optically thick lines to the
total line force as a function of radius. The division between
optically thick and optically thin lines was selected at the unity radial
Sobolev optical depth. The figure was plotted for three selected models. The
optical depths and the line force were calculated in the Sobolev approximation.}
\label{alfa}
\end{figure}

Both optically thick and optically thin lines contribute to the radiative
driving. This is illustrated in Fig.~\ref{alfa}, where we plot the ratio of the
optically thick line force to the total line force. This ratio also corresponds
to the $\alpha$ parameter and it should be constant for line strength
distribution function with uniform shape. This is true in some parts of
the figure for the low-density 100-40 model. However, the ratio significantly varies
with radius for higher density model 150-40 as a result of curvature of line
strength distribution function. The line driving is dominated by optically thick
lines close to the star (for low velocities). The
density decreases with increasing velocity, thereby reducing the number of optically thick lines. As a result,
optically thin lines become more important at larger radii. Abrupt changes of
the ratio appear owing to the change of the line strength from optically thick to
optically thin for the unity optical depth. There are more optically thick lines
in the model 150-40 (mostly of iron), consequently the variations are smoother
in this model.

\end{document}